# High-Frequency Tunable Grounded and Floating Incremental-Decremental Meminductor Emulators and Application

Garima Shukla*, Pratik Kumar†, Sajal K. Paul*+

+ sajalkpaul@rediffmail.com

*Department of Electronics Engineering, Indian Institute of Technology (ISM), Dhanbad, India, 826004

† Centre for Nano Science and Engineering, Indian Institute of Science, Bangalore, India, 560012

**Abstract**

This paper proposes a new design for realizing grounded and floating meminductor emulators built with two operational transconductance amplifiers (OTAs) and two second-generation current conveyors. The proposed grounded and floating emulators claim that the circuits are much simpler in design and can be utilized in incremental and decremental topologies. The proposed circuits' performance has been verified with Cadence Virtuoso Spectre using standard CMOS 180nm technology. Furthermore, the layout of the proposed circuits has been designed, and post-layout simulations have been performed. The non-ideal and Monte Carlo analyses have been carried out in detail. This paper also proposes the application of a meminductor as an Amplitude Modulator (AM). Moreover, the experimental results are presented to verify the theoretical and simulation analyses of proposed meminductor emulator circuits.

## 1. Introduction

Resistor, inductor, and capacitor were three traditional fundamental basic electrical elements; now, the memristor represents the fourth fundamental element. Chua postulated the memristor in 1971 [1] as the fourth basic electrical element. In 1980 this postulation was then generalized to an infinite variety of basic circuit elements [2] and can be generalized into elements quadrangle. It was highlighted only after 2008 when HP fabricated a memristor based on thin-film $TiO_2$. From then onward, there has been a boom of research in this field. Chua's circuit element quadrangle was then extended to propose higher-order elements (which require two or more than two Cs ), such as memcapacitors (MCs) and meminductors (MLs). The meminductor provides a relationship between the charge q and the time integral of flux ρ. Unlike capacitors and inductors, meminductors can store information for a long without power because of their non-volatile property. Although the device is still a theoretical concept, some device-level memelements (only memristor) have been fabricated [3], and emulators are essential to analyze the characteristics and study their applications. The research on solid-state memelements is yet to mature completely, especially for MCs and MLs. Solid-state MCs have not been commercialized, and there has been no information on solid-state MLs. Some models, though directly labeled as "memristors" or "memcapacitors," are essentially practical MR emulators [3]. Therefore, a substantial number of circuit implementations have been proposed through the use of emulators. In [4], a relationship on the doubly periodic table of circuit elements, also called the four elements torus, is given in correlation with the basic circuit element quadrangle of all four basic electrical elements. Furthermore, an extension of the memristive system to capacitive and inductive elements whose properties depend on the state and history of the system is presented in [5]. Physical characteristics analysis of these memory-based elements and mathematical examples for memristor, meminductor, and memcapacitors are presented in [6, 7].

Several circuits for emulating memristor-less meminductors are proposed in [9-23], while meminductors formed using mutators are proposed in [25-28]. In [8], a memristor-less current and voltage-controlled meminductor emulator are reported using a second-generation current conveyor (CCII), adder, multiplier, and several passive components in the count. A charge-controlled meminductor emulator using an inductor, op-amps, multiplier, transistors, and several other passive components is reported in [9]. In 2014, a practical implementation of the meminductor using many active and passive components was reported [10]. It consists of four current feedback operational amplifiers (CFOAs), one buffer, two op-amps, one multiplier, and some passive components making the circuit quite complex and bulky. A flux-controlled meminductor is reported [11] but consists of many active blocks and passive components. The meminductor reported in [12] is based on six op-amps and one multiplier, whereas the design in [13] employs three CFOAs, one op-amp, one operational transconductance amplifier (OTA), and one multiplier. The design reported in [14] is based on two voltage differencing transconductance amplifiers (VDTAs) and one multiplier. In 2017, a much simpler circuit for emulating a meminductor was reported using multioutput OTA [15], but it uses an inductor and has a low frequency of operation. All the reported circuits [8-14] are complex as they employ multiplier along with an excessive number of active blocks, and [15] have a shallow frequency of operation, which very much limits the practical use and, further, it realizes only grounded meminductor. The meminductor design in [16] is based on three OTAs and two capacitors. The design reported in [17] is based on two OTAs and one differential voltage current conveyor (DVCC), and the design in [18] is based on one OTA and one VDTA. The topology in [19] reports a memeinductor employing two OTAs and one current differencing buffered amplifier (CDBA), whereas [20] is based on two OTAs and one current differencing transconductance amplifier (CDTA). Meminductor design in [21] employs two CCIIs and one OTA, whereas design [22] is based on two VDTAs. Moreover, [23] reports a design based on one modified voltage differencing current conveyor (MVDCC) and one OTA. However, all the designs reported in [8-12, 15-18, 20-23] realize only one type of meminductance emulator, i.e., the grounded or floating meminductor; only [19] realizes both grounded and floating meminductance emulator. Furthermore, the designs reported in [21-23] realize only one type of meminductance emulator and possess a low frequency of operation. Moreover, [22, 23] realize only incremental type of configuration.

Another method of emulating meminductors is with mutators proposed in [24-28]. Mutators simulating second-order elements using their inherent relationship are reported in [24, 25] and represent the simplified meminductor emulator using the multiplier approach, but the memristor used here is bulky, complex, and has a low operating frequency. A mutator based on one CCII and three op-amps is reported in [26]. A universal mutator using many active and passive components is also available [27]. Moreover, a mutator circuit based on two current buffered transconductance amplifiers (CBTAs) and one multiplier is reported in [28]. Apart from these mutators, the PSpice model of meminductor and its nonlinear model with its study on device parameter variations are available in [29, 30]. A detailed composite behavior in series and parallel meminductor topologies is reported [31]. The applications of meminductors in chaotic oscillators and their dynamic studies are reported in [32-34], whereas application as a low-power filter design is available in [35].

This paper proposes a meminductor emulator built with active blocks consisting of two OTAs and two-second generation current conveyors. The proposed memristor emulators possess the following important features: (i) simple circuitry with no requirement of multiplier, (ii) grounded and floating configuration from the same topology, (iii) option for both incremental and decremental configurations to increase the range of values of meminductance ( the value of meminductance can be increased and decreased from its base value in incremental and decremental types of topology respectively), and also application flexibility, (iv) high-frequency range of operation, (v) electronic control of meminductance value in addition to the control by frequency and amplitude of the applied voltage signal across emulator.

## *2.* Employed Analog Building Blocks and General Meminductor Model

### 2.1 Operational transconductance amplifier (OTA), Second generation Current Conveyor (CCII) and Second-generation current controlled Current Conveyor (CCCII)

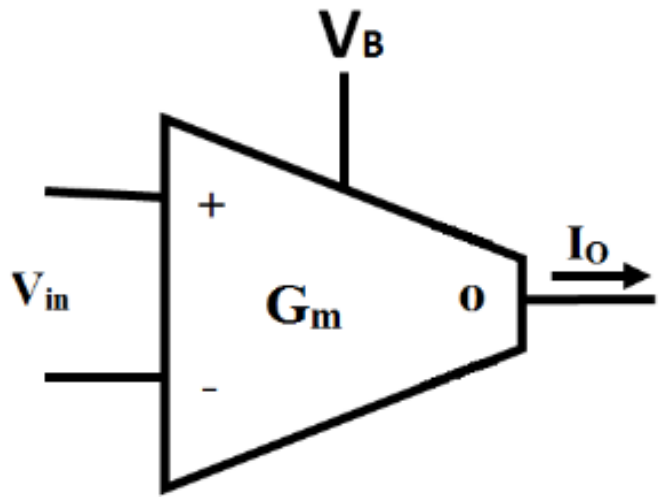

(a)

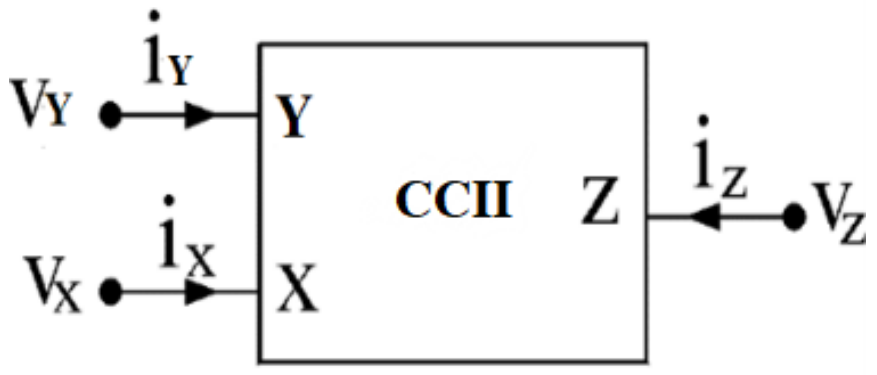

(b)

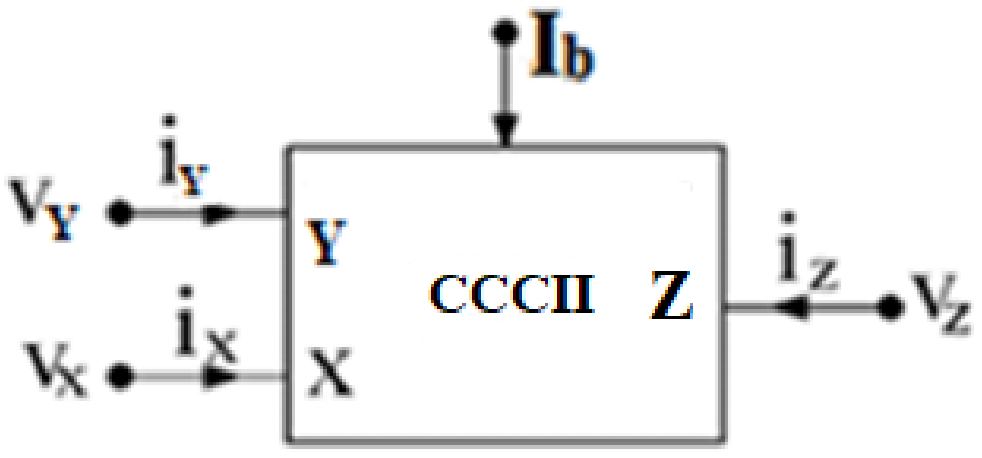

(c)

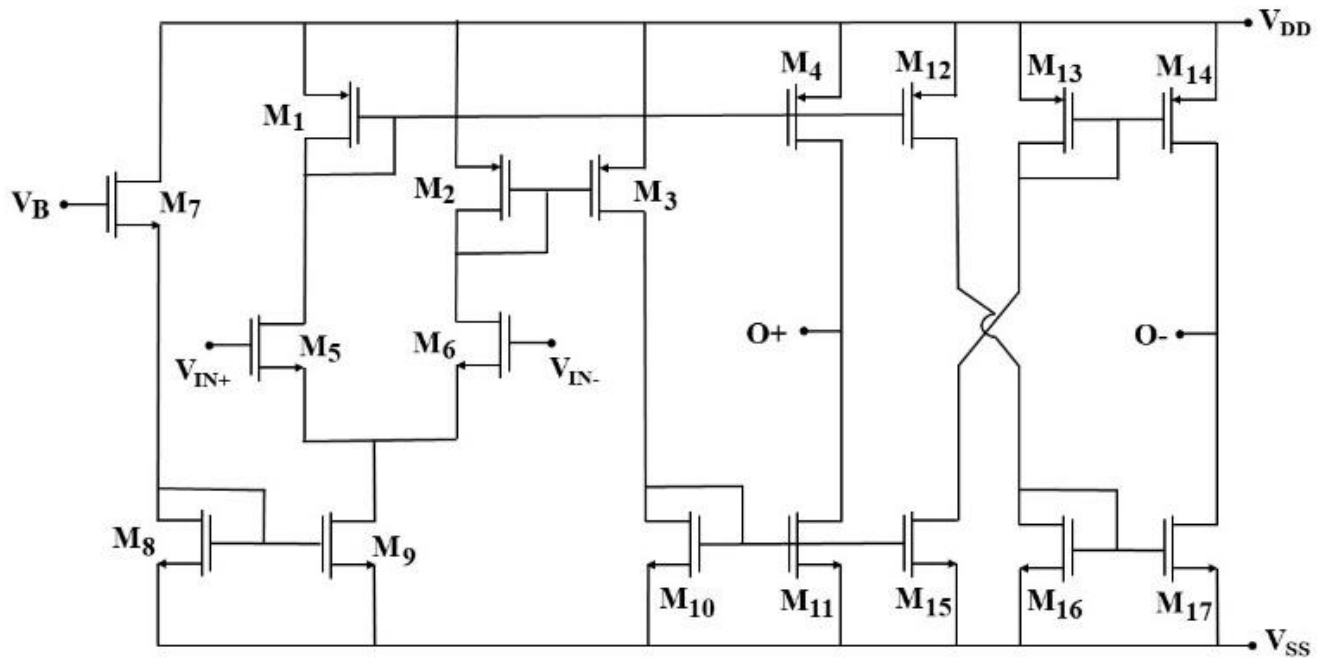

(d)

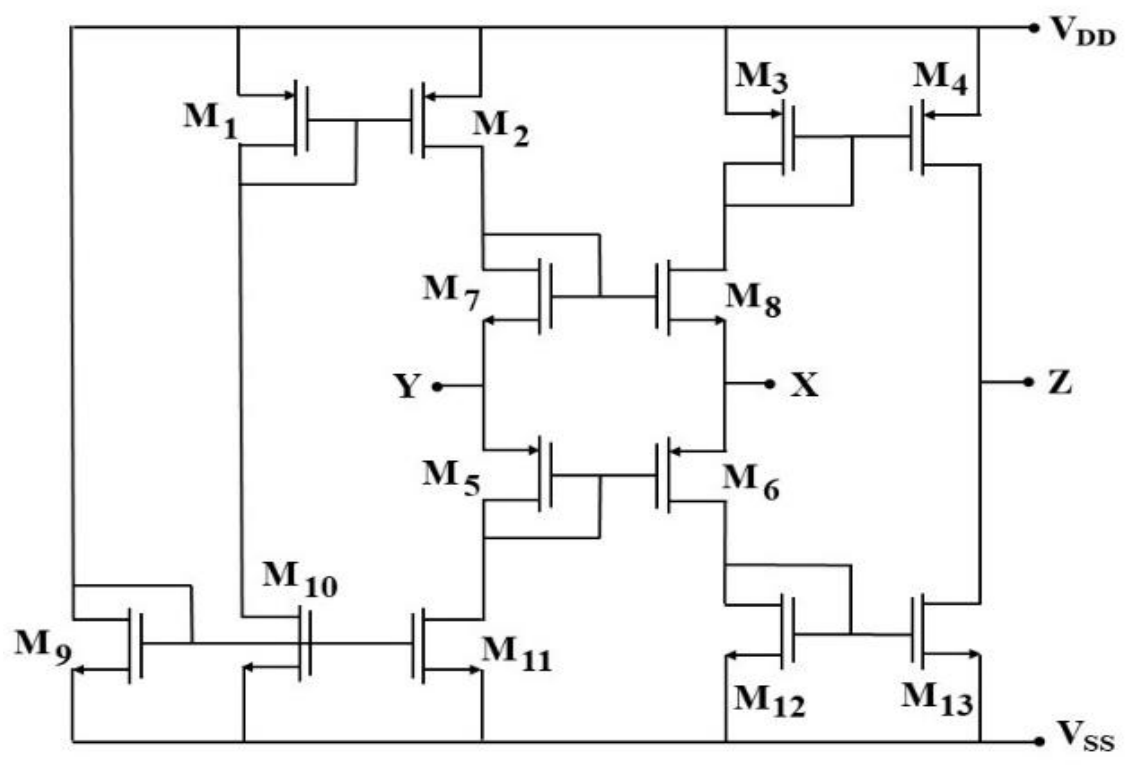

(e)

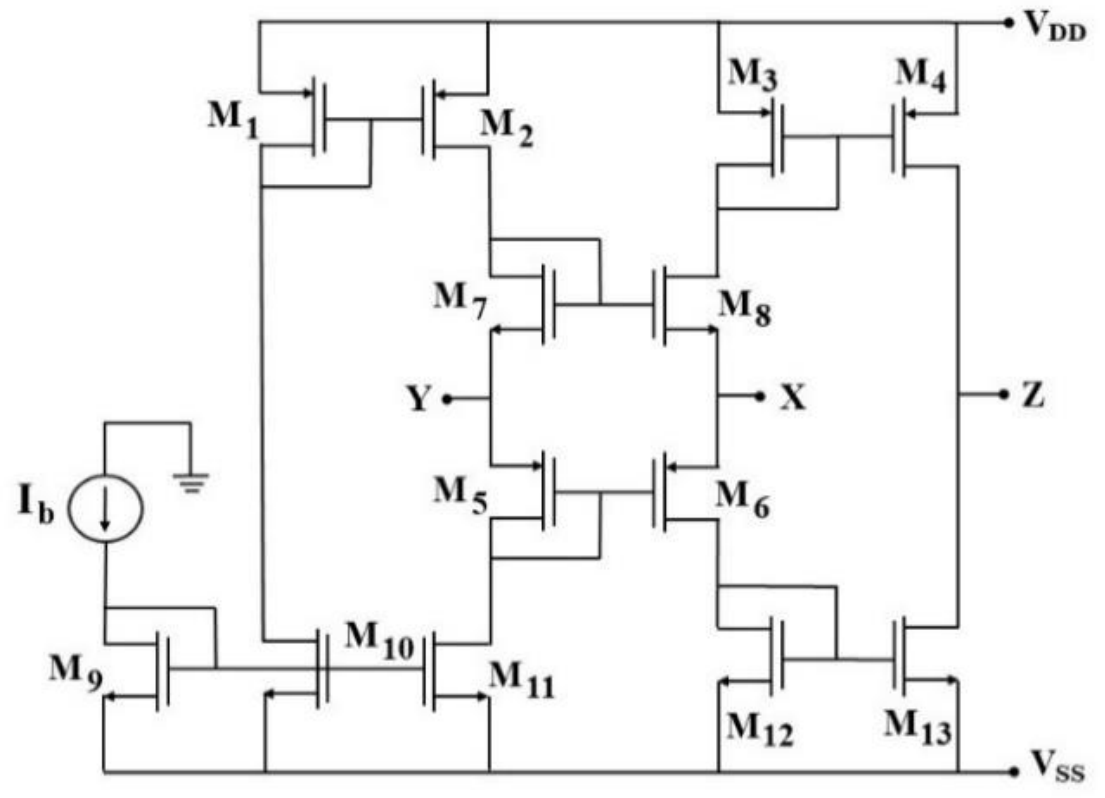

(f)

**Figure. 1.** Symbolic representations of (a) OTA, (b) CCII, (c) CCCII; CMOS implementations of; (d) OTA, (e) CCII, (f) CCCII.

OTA, CCII, and CCCII circuit symbols are shown in Figure. 1(a)-(c). CMOS implementation of OTA, CCII and CCCII are presented in Figure. 1(d), 1(e) and 1(f) respectively. $V_B$ controls the OTA's transconductance gain ($G_m$), while $I_b$ controls the internal resistance $R_X$ of CCCII, making the circuits electronically tunable.

The port relationships of OTA are expressed as:

$I_{O\pm} = \pm G_m V_{in}$ ,

$V_{in+} - V_{in-}$ =differential input = $V_{in}$

Where $G_m$ is the transconductance of OTA. The routine analysis results in the following

expression for $G_m$:

$$G_m = \frac{k}{\sqrt{2}}(V_B - V_{ss} - 2V_{th}), \quad (1)$$

Here, k is a parameter of the MOS device given by:

$$k = \mu_n C_{ox} \frac{W}{L}$$

The W, L, $\mu_n$, $C_{ox}$, and $V_{th}$ are, respectively, channel width, length, the mobility of the carrier, capacitance per unit area, and the threshold voltage of MOS.

The port relationship CCII± is given by:

$$\begin{bmatrix} I_Y \\ V_X \\ I_Z \end{bmatrix} = \begin{bmatrix} 0 & 0 & 0 \\ 1 & 0 & 0 \\ 0 & \pm 1 & 0 \end{bmatrix} \begin{bmatrix} V_Y \\ I_X \\ V_Z \end{bmatrix}$$

Similarly, port relationship CCCII± is given by:

$$\begin{bmatrix} I_Y \\ V_X \\ I_Z \end{bmatrix} = \begin{bmatrix} 0 & 0 & 0 \\ 1 & R_X & 0 \\ 0 & \pm 1 & 0 \end{bmatrix} \begin{bmatrix} V_Y \\ I_X \\ V_Z \end{bmatrix};$$

where, $R_X = \frac{1}{\sqrt{2 I_b C_{ox}}\left(\sqrt{\frac{\mu_p W_p}{L_p}} + \sqrt{\frac{\mu_n W_n}{L_n}}\right)}$

Figure. 2(a-b) shows the OTA, CCII, and CCCII frequency responses. They result in a bandwidth of 80 MHz for OTA and 1 GHz and 800 MHz for CCII and CCCII, respectively.

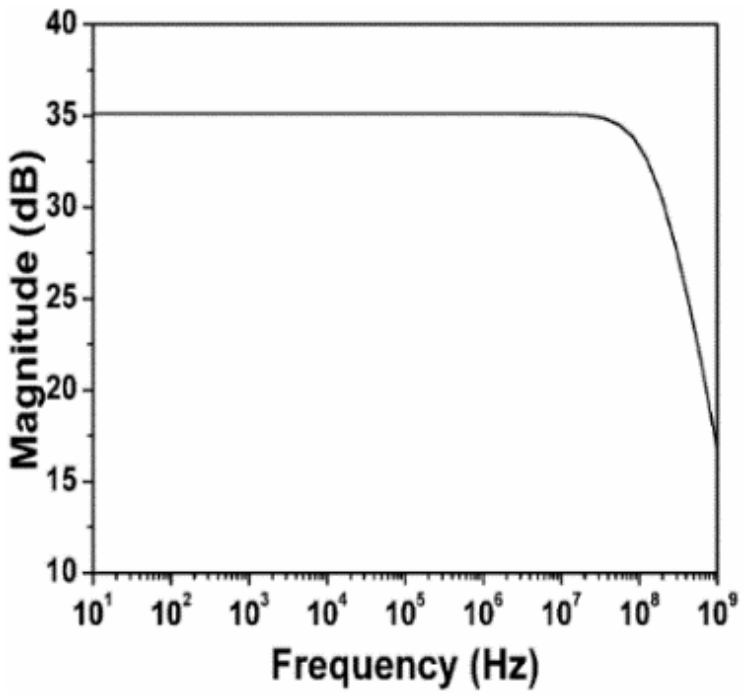

(a)

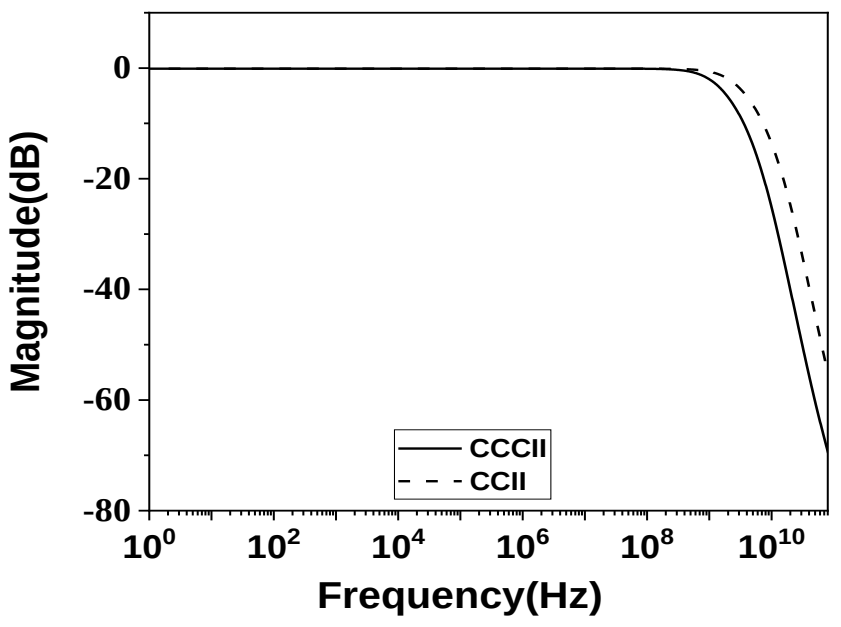

(b)

**Figure. 2.** Frequency response of (a) OTA, (b) CCII and CCCII.

## 2.2 Basic model of a meminductor emulator

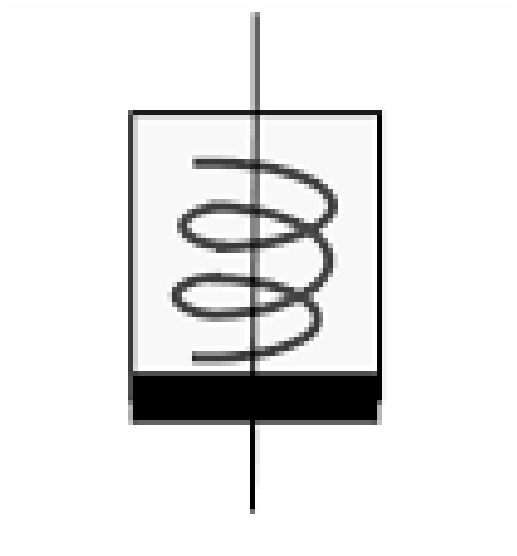

**Figure. 3.** Symbolic representation of meminductor.

Symbolic representation of meminductor is shown in Figure. 3. Meminductor is a mem-element with three constitutive variables as; $\phi(t)$, $\rho(t)$, and I(t), where $\phi(t)$ is the flux, which is defined as the integral of input voltage i.e.

$$\Phi(t) = \int V_{in}(t)dt \tag{2(a)}$$

$\rho(t)$ is the integral of $\phi(t)$, i.e.

$$\rho(t) = \int \Phi(t)dt \tag{2(b)}$$

The relation between input current I(t) and $\phi(t)$ of meminductor is defined as;

$$\frac{I(t)}{\phi(t)} = {L_M}^{-1} \tag{3(a)}$$

${L_M}^{-1}$ is inverse meminductance and the general representation of flux controlled meminductor having an initial value of inverse meminductance given by 'm' and decremental or incremental product term given by n is expressed as [14, 18];

$$\frac{I(t)}{\phi(t)} = m \pm n\rho(t) \qquad 3(b)$$

or, $I(t) = (m \pm n\rho(t))\phi(t)$

It can be inferred from (3b) that a meminductor model contains 'm' as a fixed term and $n\rho(t)$ as a time varying term.

## 3. Proposed Grounded and Floating Meminductor Emulator Circuit

Schematic diagrams of the proposed grounded and floating meminductor emulators are shown in Fig. 4 and Fig. 5, respectively. The Incremental and decremental nature of the meminductor can be configured by switching mechanism among pins M, N, O, and P of circuits as given in Table 1. These mechanisms apply to both grounded and floating meminductor emulators.

**Table 1.** Connection topology for pins M, N, O, and P for two modes of operations.

| S. No. | Switch Connections | Mode of operation |
|---|---|---|
| **1** | M-N; O-P | Incremental |
| **2** | M-P; O-N | Decremental |

*3.1.Grounded meminductor emulator*

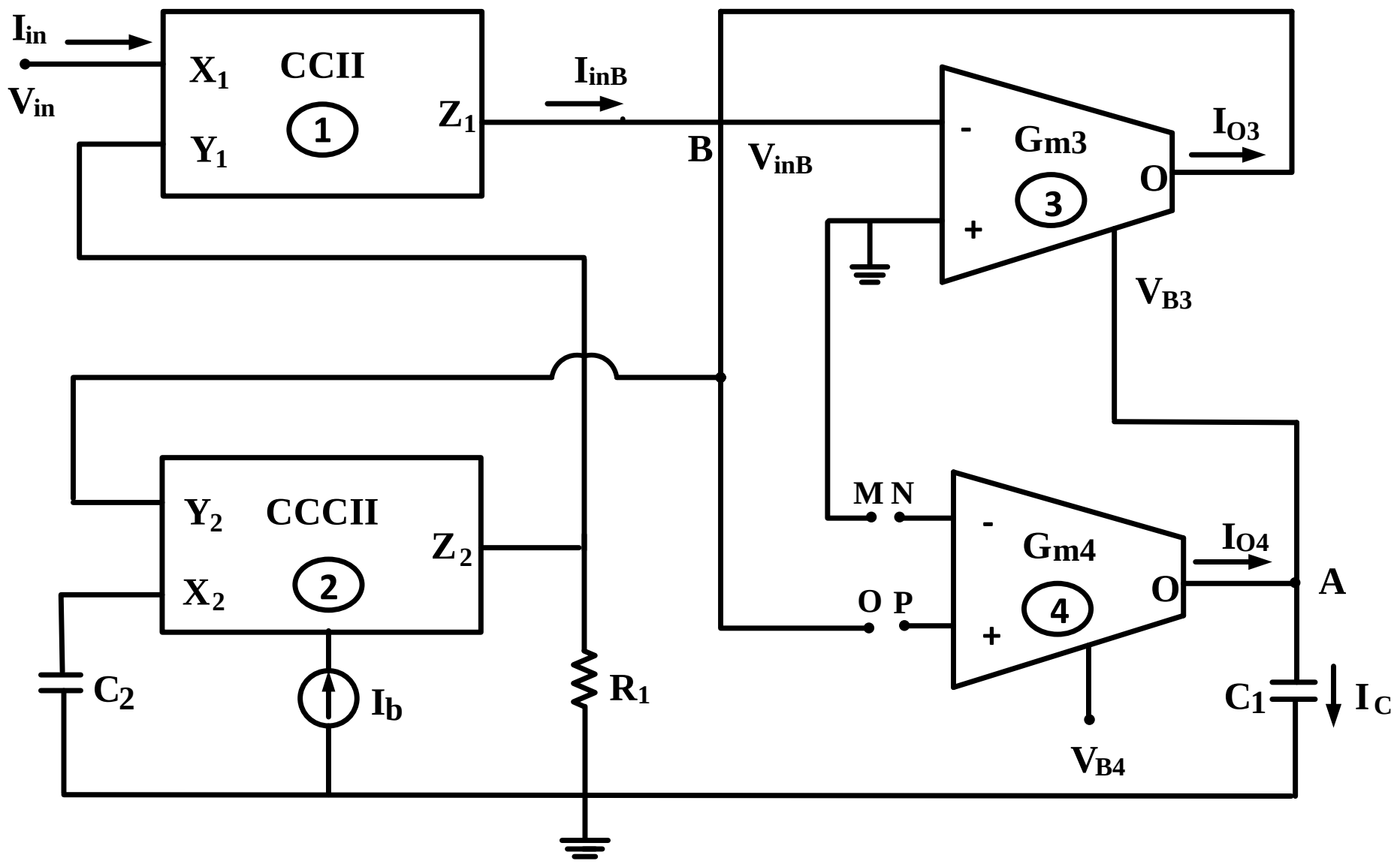

**Fig. 4.** Schematic diagram of grounded meminductor emulator

The proposed grounded meminductor emulator is shown in Fig. 4. Considering the incremental type of meminductor emulator, i.e., pins M, N and O, P are interconnected, the input current $I_{in}$ is obtained as:

$$I_{in}(t) = I_{X1} = I_{Z1} \quad \text{(using port relationship of CCII)}$$

Again, $$I_{in}(t) = -I_{inB} \tag{2}$$

Similarly, $V_{Y2} = V_{X2} - I_{X2}R_{X2}$ (by the port relationship of CCCII)

$$V_{Y2} = V_{inB} = -\frac{I_{X2}}{sC_2} - I_{X2}R_{X2} = -I_{X2}(\frac{1}{sC_2} + R_{X\,2}) \tag{3}$$

Again, $I_{X2} = I_{Z2} = \frac{-V_{Y1}}{R_1} = \frac{-V_{in}}{R_1}$ , since $(V_{Y1} = V_{X1} = V_{in})$ (4)

Substituting (4) in (3)

$$V_{in} = V_{inB}\left(\frac{sR_1C_2}{1+sC_2R_{X2}}\right) \tag{5}$$

Dividing (5) by (2)

$$\frac{V_{in}}{I_{in}} = -\frac{V_{inB}}{I_{inB}}\frac{sR_1C_2}{1+sC_2R_{X2}} \tag{6}$$

Bias voltage $V_{B3}$ is given by

$$V_{B3} = \frac{1}{C_1}\int I_C(t)dt = \frac{G_{m4}}{C_1}\int V_{inB}(t)dt = \frac{G_{m4}}{C_1}\left(\frac{1+sC_2R_{X2}}{SR_1C_2}\right)\int V_{in}(t)dt$$

$$= \frac{G_{m4}}{C_1}\left(\frac{1+sC_2R_{X2}}{sR_1C_2}\right)\phi_{in} \tag{7}$$

Where $\phi_{in}$ is the total flux obtained by the meminductor, and it is expressed as

$$\varphi_{in} = \int V_{in}(t)dt = \frac{V_{in}}{s} \tag{8}$$

Substituting (7) into (1), the transconductance $G_{m3}$ is obtained as

$$G_{m3} = \frac{k}{\sqrt{2}}(V_{B3} - V_{ss} - 2V_{th}) = \frac{k}{\sqrt{2}}\left(\frac{G_{m4}(1+sC_2R_{X2})\phi_{in}}{sC_1R_1C_2} - V_{ss} - 2V_{th}\right) \tag{9}$$

Also, $$G_{m3} = \frac{-I_{O3}}{V_{inB}} = \frac{I_{inB}}{V_{inB}} \tag{10}$$

Using (10) and (9) results in an expression

$$\frac{I_{inB}}{V_{inB}} = \frac{k}{\sqrt{2}}\left(\frac{G_{m4}(1+sC_2R_{X2})\phi_{in}}{sC_1R_1C_2} - V_{ss} - 2V_{th}\right) \tag{11}$$

On substituting (11) in (6) and by incorporating the relation of (8), meminductance ($L_M$) of the proposed grounded incremental meminductor emulator is obtained as

$$L_M = \frac{\phi_{in}}{I_{in}} = \frac{R_1C_2}{\frac{k}{\sqrt{2}}(V_{ss}+2V_{th}) - \frac{k}{\sqrt{2}}\left(\frac{G_{m4}(1+sC_2R_{X2})\phi_{in}}{sC_1R_1C_2}\right)}\frac{1}{1+sC_2R_{X2}} \tag{12}$$

Similarly, the change of switch connections to w-z and y-x changes the polarity of the time-variant part of the meminductance of (12), resulting in a decremental type meminductance as

$$L_M = \frac{R_1C_2}{\frac{k}{\sqrt{2}}(V_{ss}+2V_{th}) + \frac{k}{\sqrt{2}}\left(\frac{G_{m4}(1+sC_2R_{X2})\phi_{in}}{sC_1R_1C_2}\right)}\frac{1}{1+sC_2R_{X2}} \tag{13}$$

Equations (12) and (13) can be combined and rewritten as

$$L_M = \frac{R_1C_2}{\frac{k}{\sqrt{2}}(V_{ss}+2V_{th}) \mp \frac{k}{\sqrt{2}}\left(\frac{G_{m4}(1+sC_2R_{X2})\phi_{in}}{sC_1R_1C_2}\right)}\frac{1}{1+sC_2R_{X2}} \tag{14}$$

If the operating frequency is much less than $\frac{1}{2\Pi C_2 R_{X2}}$ where $R_{X2}$ is the resistance at input port X of CCCII and is usually $R_{X2}$ kept low [39], then we can write $(1 + sC_2R_{X2}) \approx 1$.

Hence, equation (14) can then be simplified in terms of the inverse of $L_M$ as:

$$L_M^{-1} = \frac{\phi_{in}}{I_{in}} \approx \frac{k}{\sqrt{2}R_1C_2}(V_{ss} + 2V_{th}) \pm \frac{k}{\sqrt{2}}\left(\frac{G_{m4}\phi_{in}}{sC_1R_1^2C_2^2}\right) \tag{15}$$

In equation (15), $G_{m4}$ can be controllable by external bias voltage $V_{B4}$, which makes the proposed circuit electronically tunable. The equations (15) represent incremental and decremental meminductance, where for a fixed operating frequency $\frac{k}{\sqrt{2}R_1C_2}(V_{ss} + 2V_{th})$ is the constant term and $\frac{k}{\sqrt{2}}\left(\frac{G_{m4}\phi_{in}}{sC_1R_1^2C_2^2}\right)$ is the time-varying term as $\phi_{in}$ is the function of the time-varying input signal. For $\phi_{in} = 0$ meminductance attains a constant value in both the topologies (incremental and decremental), where for the operator±, the + is for decremental and – is for incremental configuration.

*3.2.Floating meminductor emulator*

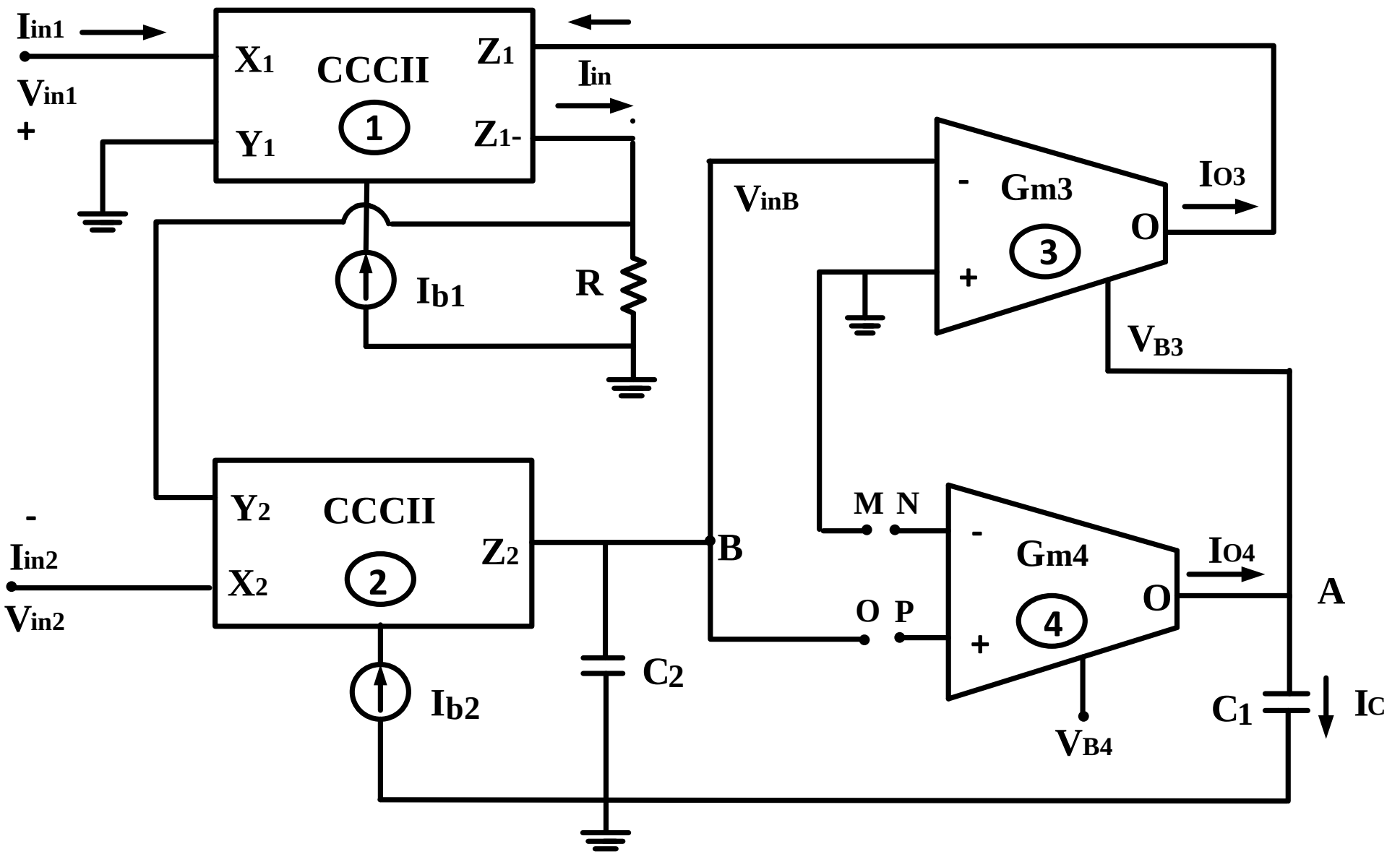

**Fig. 5.** Schematic diagram of floating meminductor emulator.

The floating meminductor emulator is shown in Fig. 5. Considering incremental type meminductor emulators, i.e., pins M, N, and pins O, P are interconnected. The currents from the port relationship are obtained as follows;

$$I_{in1} = I_{in} = I_{X1} = I_{Z1} = -I_{Z1-}, \quad (16)$$

And $$V_{in1} = V_{X1} = V_{Y1} + I_{X1}R_{X1} \quad (17)$$

Hence, $$V_{in1} = I_{in}R_{X1} \quad [\text{As } V_{in1}=0] \quad (18)$$

Again, $$V_{Z1-} = V_{Y2} = I_{Z1-}R = \frac{V_{in1}}{R_{X1}} * R \quad (19)$$

For $R = R_{X1}$, we get from (19), $V_{Y2} = V_{in1}$ (20)

Further, $$V_{in2} = V_{X2} = V_{Y2} + I_{X2}R_{X2} \quad [\text{from port relation}] \quad (21)$$

Hence from (20) and (21), $$V_{in2} = V_{in1} + I_{X2}R_{X2} \quad (22)$$

As at node B, $$I_{X2} = I_{Z2} = -sC_2V_{inB} \quad (23)$$

Hence from (22), $$V_{in2} = V_{in1} - sC_2V_{inB}R_{X2} \quad (24)$$

Or $$V_{in1} - V_{in2} = V_{in} = sC_2V_{inB}R_{X2} \quad (25)$$

From (25), $V_{in} = sR_{X2}C_2V_{inB}$ or $V_{inB} = \left(\frac{V_{in}}{sR_{X2}C_2}\right)$ (26)

Also $I_{in} = I_{O3}$ (27)

Dividing (26) by (27), we get;

$$\frac{V_{in}}{I_{in}} = \frac{V_{inB}}{I_{O3}} sR_{X2}C_2 \tag{28}$$

Further, bias voltage $V_{B3}$ using (26) results in the following;

$$V_{B3} = \frac{1}{C_1}\int I_C(t)dt = \frac{G_{m4}}{C_1}\int V_{inB}(t)dt = \frac{G_{m4}}{C_1}\left(\frac{1}{sR_{X2}C_2}\right)\int V_{in}(t)\,dt$$

$$= \frac{G_{m4}}{C_1}\left(\frac{1}{sR_{X2}C_2}\right)\phi_{in} \tag{29}$$

Substituting (29) into (1), transconductance $G_{m3}$ is obtained as;

$$G_{m3} = \frac{k}{\sqrt{2}}(V_{B3} - V_{ss} - 2V_{th}) = \frac{k}{\sqrt{2}}\left(\frac{G_{m4}\phi_{in}}{sC_1R_{X2}C_2} - V_{ss} - 2V_{th}\right) \tag{30}$$

Further from Fig. 5, $G_{m3} = -\frac{I_{O3}}{V_{inB}}$

So (30) becomes :

$$\frac{I_{O3}}{V_{inB}} = -\frac{k}{\sqrt{2}}\left(\frac{G_{m4}\phi_{in}}{sC_1R_{X2}C_2} - V_{ss} - 2V_{th}\right) = \frac{k}{\sqrt{2}}\left(V_{ss} + 2V_{th} - \frac{G_{m4}\phi_{in}}{sC_1R_{X2}C_2}\right) \tag{31}$$

Substituting (31) in (28) gives,

$$\frac{V_{in}}{I_{in}} = \frac{sR_{X2}C_2}{\frac{k}{\sqrt{2}}(V_{ss}+2V_{th}) - \frac{k}{\sqrt{2}}\left(\frac{G_{m4}\phi_{in}}{sC_1R_{X2}C_2}\right)}$$

Hence, the meminductance of the proposed grounded incremental meminductor emulator is obtained as;

$$L_M = \frac{R_{X2}C_2}{\frac{k}{\sqrt{2}}(V_{ss}+2V_{th}) - \frac{k}{\sqrt{2}}\left(\frac{G_{m4}\phi_{in}}{sC_1R_{X2}C_2}\right)} \tag{32}$$

Similarly, the change of switch connections to M-P and O-N changes the polarity of the time-variant part of meminductance of (24), resulting in a decremental type meminductance as;

$$L_M = \frac{R_{X2}C_2}{\frac{k}{\sqrt{2}}(V_{ss} + 2V_{th}) + \frac{k}{\sqrt{2}}\left(\frac{G_{m4}\phi_{in}}{sC_1R_{X2}C_2}\right)} \tag{33}$$

So equations (32) and (33) can be combined and rewritten as ;

$$L_M = \frac{\phi_{in}}{I_{in}} = \frac{R_{X2}C_2}{\frac{k}{\sqrt{2}}(V_{ss}+2V_{th}) \pm \frac{k}{\sqrt{2}}\left(\frac{G_{m4}\phi_{in}}{sC_1R_{X2}C_2}\right)}$$

$$L_M{}^{-1} = \frac{I_{in}}{\phi_{in}} = \frac{k}{\sqrt{2}R_{X2}C_2}(V_{ss} + 2V_{th}) \pm \frac{k}{\sqrt{2}}\left(\frac{G_{m4}\phi_{in}}{sC_1R_{X2}{}^2C_2{}^2}\right) \quad (34)$$

In equation (34), $G_{m4}$ can be controllable by external bias voltage $V_{B4}$, and $R_{X2}$ is controlled by external current $I_{b2}$, which makes the proposed circuit electronically tunable. In the inverse of the meminductance equation, the term $\frac{k}{\sqrt{2}R_{X2}C_2}(V_{ss} + 2V_{th})$ is constant, and $\frac{k}{\sqrt{2}}\left(\frac{G_{m4}\phi_{in}}{SC_1R_{X2}{}^2C_2{}^2}\right)$ is the time-varying term as $\phi_{in}$ is the function of the time-varying input signal. For $\varphi_{in} = 0$, meminductance attains a constant value in both the topologies (incremental and decremental), where for the operator ±, the + is for decremental and – is for incremental configuration.

## 4. Comparison of meminductor emulators

A comparison of available meminductor emulators is given in Table 2. It is observed that most emulators use many analog building blocks for implementation and have frequency limitations. The emulator designs reported in [8-12, 26, 27] use many active building blocks and a large number of passive components and do not possess electronic tunability. Only the configurations [13, 14, 19] possess both grounded and floating types of meminductors, hence finding flexibility in applications. Moreover, the topologies [8-12, 15-18, 21, 25-27] are only grounded type, and [20, 23, 23, 28] are floating type of meminductors. Further, all these designs [8-14, 26-28] employ a multiplier in the configuration, which is undesirable as it increases circuit complexity. The design in [15] uses a floating inductor in the configuration, resulting only in grounded types of meminductor. Furthermore, the designs [8-13, 15, 21-26] exhibit low frequency of operation (few Hz to few kHz range). Meminductors [8-13, 22, 23, 28] can be operated only in incremental configuration.

The proposed work presents both grounded and floating types of meminductor realizations using simple basic blocks, two CCII/CCCII, and two OTAs with only two capacitors and one resistor. All the passive elements in both of the proposed meminductor circuits are grounded.

Moreover, both the incremental and decremental properties are present in the proposed emulators. Further, the proposed grounded and floating meminductors are valid for frequencies of 1 MHz and 10 MHz, respectively. An important feature of the proposed meminductor emulator is its ability to control the meminductance value by controlling the transconductance, $G_{m4}$ with the bias voltage, $V_{B4;}$ hence both the emulator circuits are electronically tunable. Power consumption of the proposed circuits is greater than that of [14, 15], however smaller than that of [21] among the available literature.

**Table 2:** Comparison of meminductor emulators.

| Ref. | Number in count and type(s) of active building blocks used | Tech. used | Passive element (C/R/L) | All grounded passive elements | M /NM based | G/ F meminductor | Electronic tunability | Max. operating frequency shown | Inc/Dec or both | P. C (W) |
|---|---|---|---|---|---|---|---|---|---|---|
| [8] | 3 CCII, 1 Multiplier, 1 Adder | CMOS | 2/3/0 | Yes | NM | G | No | 20 Hz | Inc | NA |
| [9] | 2 Op-Amp, 2 Current Mirrors,1 Buffer, 1 Multiplier | CMOS | 2/2/1 | No | NM | G | No | 300 Hz | Inc | NA |
| [10] | 4 CFOAs, 1 Buffer, 2 Op-Amp, 1 Multiplier | CMOS | 2/6/0 | No | NM | G | No | 36.9 Hz | Inc | NA |
| [11] | 5 Op-Amp, 1 Multiplier | CMOS | 2/10 | No | NM | G | No | 70 Hz | Inc | NA |
| [12] | 6 Op-Amps, 1 Multiplier | BJT | 2/13/0 | No | NM | G | No | 300 Hz | Inc | NA |
| [13] | 1 OTA, 3 CFOA, 1 Op-Amp, 1 Multiplier | BJT | 2/8/0 | No | NM | Both (G+F) | Yes | 5 kHz for both | Inc | NA |
| [14] | 2 VDTAs, 1 Multiplier | CMOS | 2/0/0 | Yes | NM | Both (G+F) | Yes | 1 MHz for both | Both | 200 µ |
| [15] | 1 MO-OTA | CMOS | 1/1/1 | No | NM | G | Yes | 500 Hz | Both | 120 µ |
| [16] | 3 OTAs | CMOS | 2/0/0 | Yes | NM | G | Yes | 10 MHz | Both | NA |
| [17] | 2 OTAs,1 DVCC | CMOS | 2/1/0 | Yes | NM | G | Yes | 10 MHz | Both | NA |
| [18] | 1 OTA, 1 VDTA | CMOS | 2/0/0 | Yes | NM | G | Yes | 3 MHz | Both | NA |
| [19] | 2 OTAs, 1 CDBA | CMOS | 2/0/0 | Yes | NM | Both (G+F) | Yes | 2 MHz for both | Both | NA |
| [20] | 2 OTAs, 1 CDTA | CMOS | 2/0/0 | Yes | NM | F | Yes | 1 MHz | Both | NA |
| [21] | 2 CCIIs, 1 OTA | CMOS | 2/2/0 | Yes | NM | G | Yes | 700 kHz | Both | 14.3m |
| [22] | 2 VDTAs | CMOS | 2/0/0 | Yes | NM | F | Yes | 700 kHz | Inc | NA |

| | | | | | | | | | | | |
|---|---|---|---|---|---|---|---|---|---|---|---|
| [23] | 1 MVDCC, 1OTA | CMOS | 2/1/0 | Yes | NM | F | Yes | 300 kHz | Inc | NA |
| [25] | 1 Microcontroller, 1 ADC, 1 Op Amp | CMOS | 1/3/0 | No | M | G | No | 8 Hz | Both | NA |
| [26] | 1 CCII, 6 Op-Amp, 1 Multiplier | CMOS | 2/10/0 | No | M | G | No | 200 Hz | Both | NA |
| [27] | 7 TOAs, 1 Op-Amp, 1 Multiplier, 3 Buffers | CMOS | 2/11/0 | No | M | G | No | 21.1 Hz | Both | NA |
| [28] | 2 CBTAs, 1 Multiplier | CMOS | 2/2/0 | No | M | F | Yes | 1 MHz | Inc | NA |
| Our Work | 2 OTAs, 1 CCCII, 1 CCII | CMOS | 2/1/0 | Yes | NM | G | Yes | 1 MHz | Both | 1.01 m |
| | 2 OTAs, 2 CCCIIs | CMOS | 2/1/0 | Yes | NM | F | Yes | 10 MHz | Both | 1.03 m |

**Note: M: Mutator; NM: Non-mutator; G: grounded; F: Floating; Inc: Incremental; Dec: Decremental; P.C: Power consumption.**

## 5. Simulations results and discussion

This section deals with verifying the hysteresis loop between flux and the current, one of the fingerprints of the meminductor. Various simulations with 180 nm CMOS technology have been performed to verify the meminductive nature of proposed emulator circuits. Supply voltages of +1.2V and -1.2 V are used for $V_{DD}$ and $V_{SS,}$ respectively, for grounded and floating meminductors. The aspect ratios of MOS transistors are given in Table 3, and they operate in the saturation region.

### *5.1. Grounded meminductor simulation result*

The grounded meminductor emulator in Figure. 4 is simulated for different frequencies. The results for the pinched hysteresis loop obtained for a sinusoidal signal of amplitude ($A_m$=140 mV) with frequencies of 100 kHz, 200 kHz, 300 kHz, 400 kHz, and 500 kHz are shown in Figure. 6. Here, the product of the capacitor ($C_2$) value and frequency (f) is kept constant (75 x $10^{-6}$ FaradHz) and $V_{B4}$=450 mV. On increasing the frequency, the pinched hysteresis loop area of Φ-I curves decreases, which satisfies (14), suggesting that the time-varying nature of the loop decreases and ultimately vanishes at a specific frequency. Figure. 7 shows the relationship between charge q(t) and ρ(t), where $\rho(t) = \int \varphi(t)dt$ and $q(t) = \int i(t)dt$. Additionally, the current, $i(t)$ flowing in a meminductor can be expressed as per [6]. The single valued function q(ρ) has been illustrated in detail in [6]. This can also be verified graphically from Figure. 7 as a single valued curve is obtained, implicitly implying that the corresponding device is a meminductor.

**Table 3.** Design Parameters for analog blocks in grounded meminductor

**OTA**

| MOS Transistors | W(μm) | L(μm) |
|---|---|---|
| $M_{1\text{-}4}$ | 12 | 0.375 |
| $M_{10}$ | 12 | 0.510 |
| $M_{5\text{-}9}$, $M_{11}$ | 12 | 0.500 |

**CCII/CCCII**

| MOS Transistors | W(μm) | L(nm) |
|---|---|---|
| $M_1$, $M_{3\text{-}13}$ | 12 | 0.500 |
| $M_2$ | 2 | 0.200 |

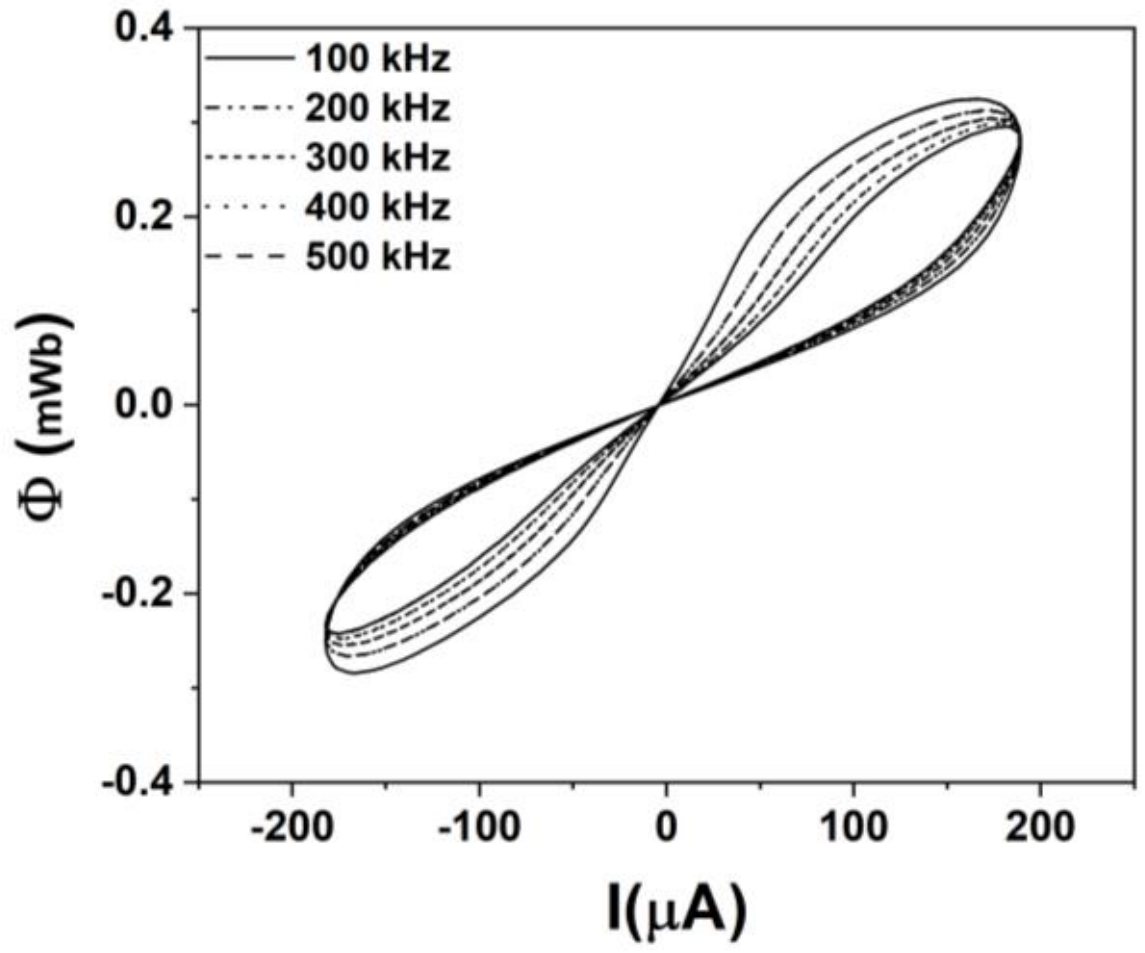

**Figure. 6.** Φ-I characteristic for grounded meminductor circuit at different operating frequencies for $A_m$ = 140 mV, $I_b$=20 μA, $V_{B4}$= 0.45 V, $C_1$= 150 pF, R=10 Ω and constant $C_2f$ =75 x $10^{-6}$

FaradHz.

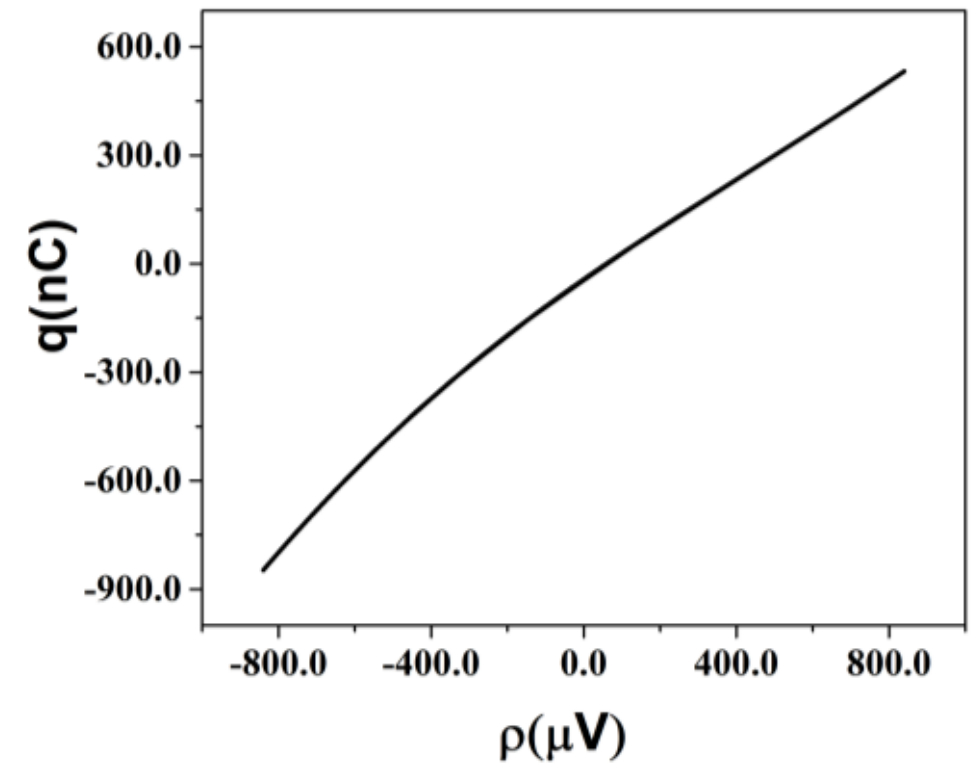

**Figure. 7.** Locus of q(t) and ρ(t) for grounded meminductor.

### *5.2. Floating meminductor simulation result*

The floating emulator's simulation results for the pinched hysteresis loop obtained for frequencies of 1 MHz, 2 MHz, 4 MHz, 6 MHz, and 8 MHz are shown in Figure. 8. Here, the product of capacitor($C_2$) value and frequency (f) is kept constant (75 x $10^{-6}$ FaradHz) with $A_m$=200 mV, $V_{B4}$=500 mV. On increasing the frequency, the pinched hysteresis loop area of Φ-I curves decreases; this validates the meminductive behavior of emulators as obtained in (34). Fig. 9 shows the relationship between charge q(t) and integral of flux, ρ(t). It shows that q(t) is a single-valued function of ρ(t). Therefore, the device current goes through a meminductor. On increasing the frequency, the pinched hysteresis loop area of Φ-I curves decreases, which satisfies (14), suggesting that the time-varying nature of the loop decreases and ultimately vanishes at a specific frequency. Figure 9 shows the relationship between charge q(t) and ρ(t). As discussed earlier for Figure 7, the curve in Figure 9 is also a single valued. Therefore, the device current flows through a meminductor.

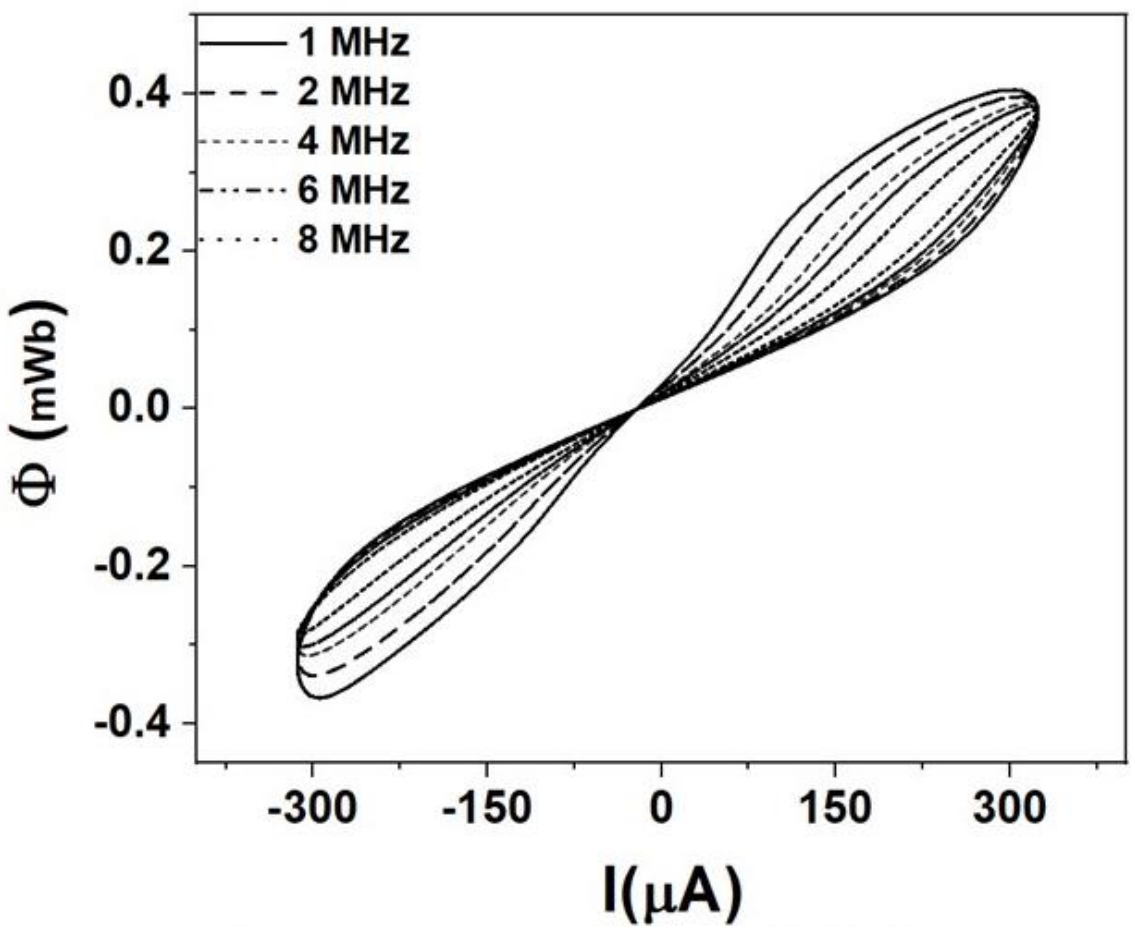

**Figure. 8.** Φ-I characteristic for floating meminductor circuit at different operating frequencies for $A_m$ = 200 mV, $I_{b2}$=20 μA, $V_{B4}$= 0.5 V, $C_1$= 150pF, R=5 Ω and constant $C_2f$ =75 x $10^{-6}$ FaradHz.

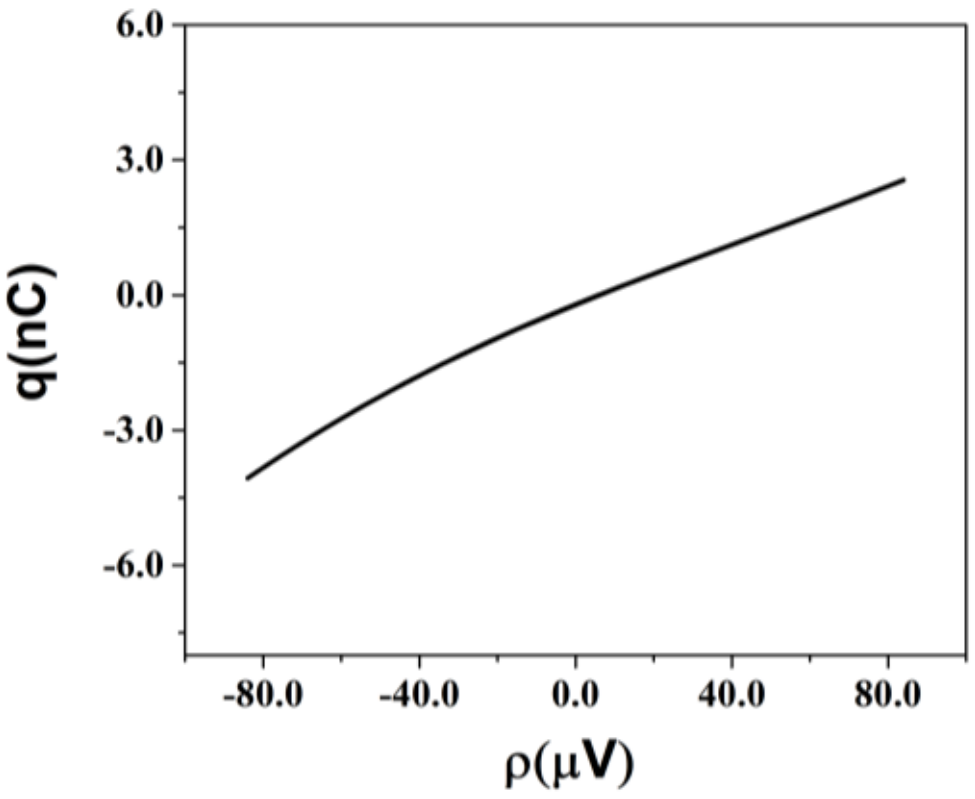

**Figure. 9.** Locus of q(t) and ρ(t) for floating meminductor.

### *5.3. Effect of variation of bias voltage ($V_{B4}$) of OTA on the pinched hysteresis loop*

It is seen in (15) and (34) that the meminductance of emulators depends on transconductance $G_{m4,}$ which is electronically tunable by the external bias voltage, $V_{B4,}$. Figure. 10(a) shows the grounded meminductor's simulation results for signal frequency of 500 kHz, capacitor value $C_1$=$C_2$=150 pF, $I_b$=20μA, and $A_m$=140 mV at different values of $V_{B4}$ (0.45 V, 0.4 V, and 0.35 V). Similarly, Figure. 10(b) shows the floating meminductor's simulation results for signal frequency of 500 kHz, capacitor value of $C_1$=375 pF, $C_2$ =150 pF, $I_{b1}$ =-48 μA, $I_{b2}$ =-20 μA and $A_m$=140 mV at different values of $V_{B4}$ (0.45V, 0.4V, and 0.35V). It is observed that the pinched hysteresis loop of Φ-I curves area increases with the increase of $V_{B4}$ as expected as per position of $G_{m4}$ in (15) and (34). It implies that the meminductance can be controlled by $V_{B4.}$

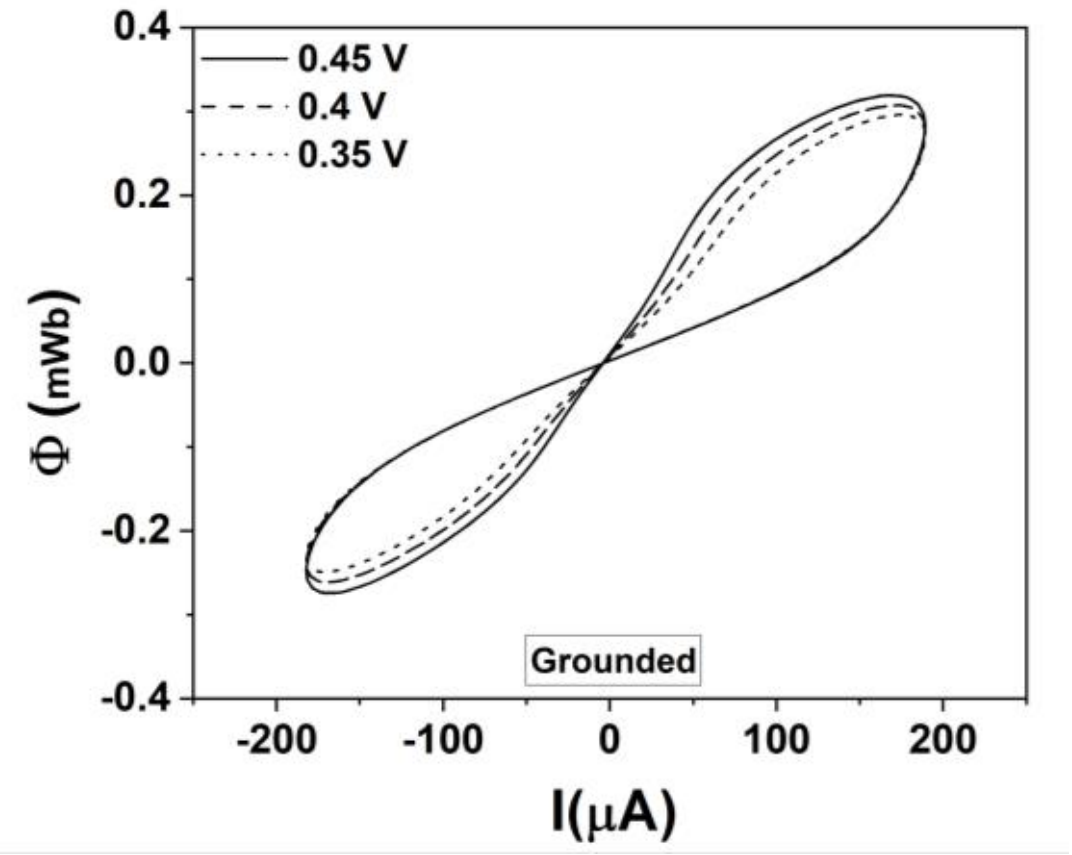

(a) Grounded

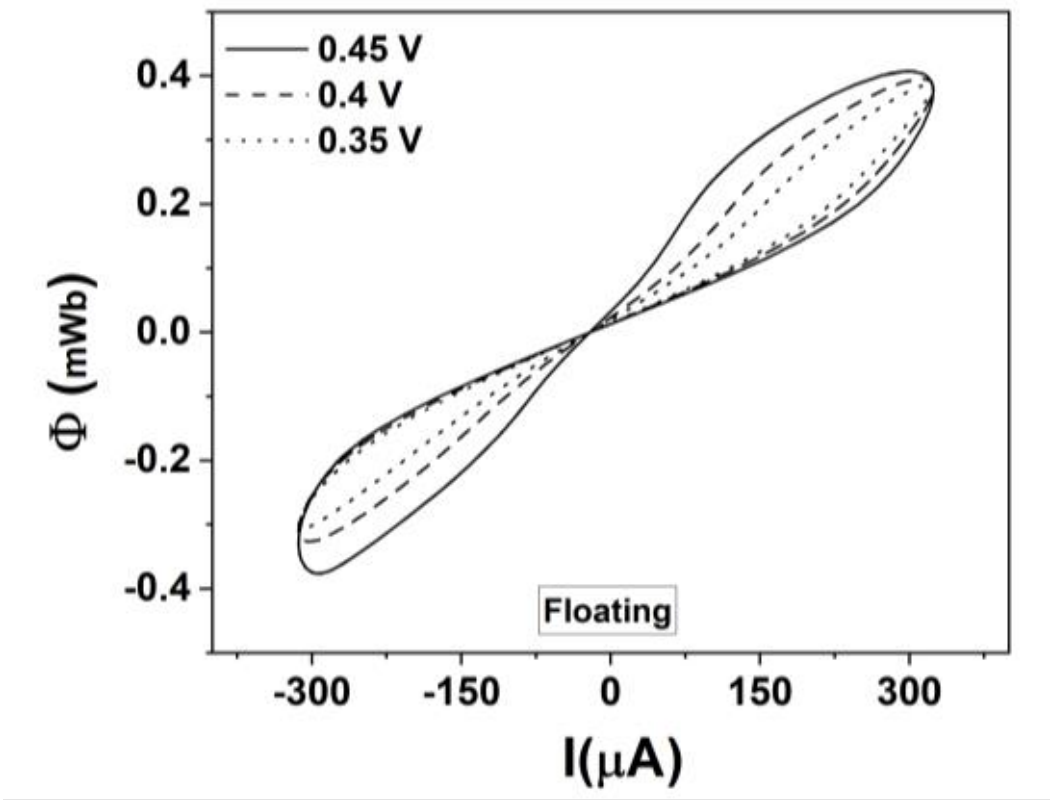

(b) Floating

**Figure. 10.** Φ-I characteristic curves obtained with sinusoidal current signal with 500 kHz, $A_m$=140 mV, $C_2$=150 pF for (a) grounded meminductor circuit with $C_1$=150 pF, $I_b$=20 μA, and different $V_{B4}$, (b) floating meminductor circuit with $C_1$=375 pF, $I_{b2}$=20 μA, and different $V_{B4}$.

***5.4 Effect of varying frequency and capacitances on the pinched hysteresis loop***

The effect on Φ-I characteristics for variation of applied signal frequency for a fixed capacitance for both grounded and floating meminductor emulators are shown in Figures. 11(a) and 11(b), respectively. Figure. 11(a) shows that as frequency increases from 400 kHz – 800 kHz for a fixed capacitance value, the hysteresis loop becomes more and more linear, which satisfies (15), suggesting the time-varying nature of the loop decreases. Figure. 11(b) shows that as the frequency increases from 800k Hz – 4 MHz for a fixed capacitance value, the area under the hysteresis loop gradually decreases as predicted by (34). Similar to the effect of frequency variation, area of hysteresis loop of meminductance changes for the variation in capacitances ($C_1$ and $C_2$) and this can also be verified by observing these parameters in (15) and (34). On varying $C_1$ and $C_2$ with a fixed frequency of 500 kHz for grounded topology and 800 kHz for floating topology, the results are obtained as shown in Figure. 11(c-f).

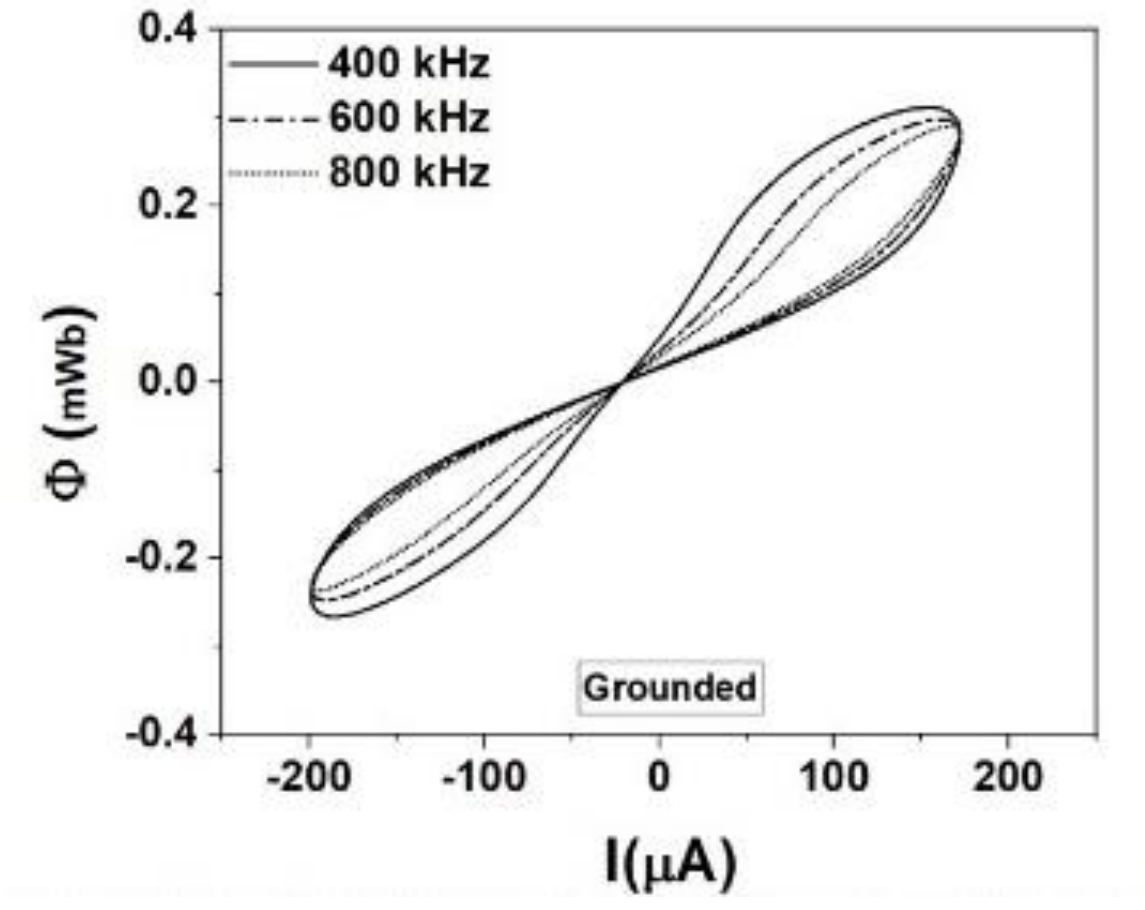

(a) Grounded

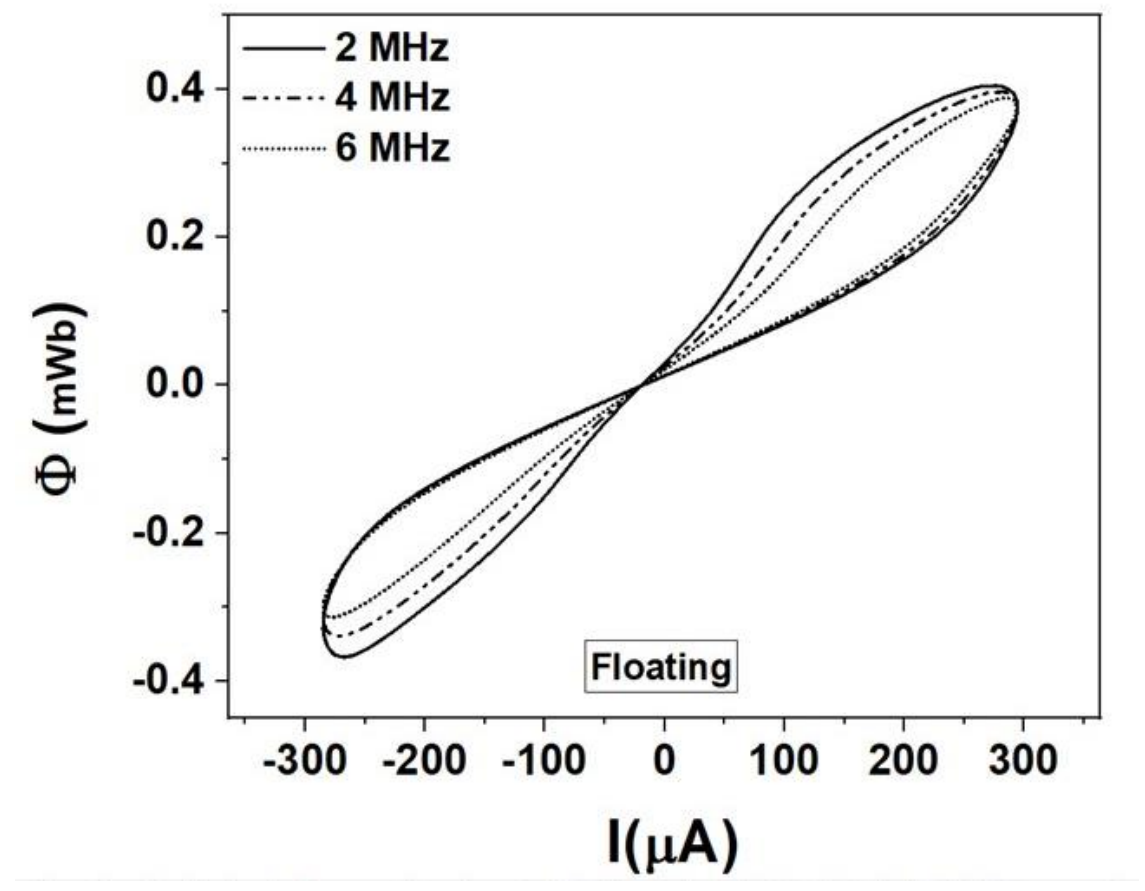

(b) Floating

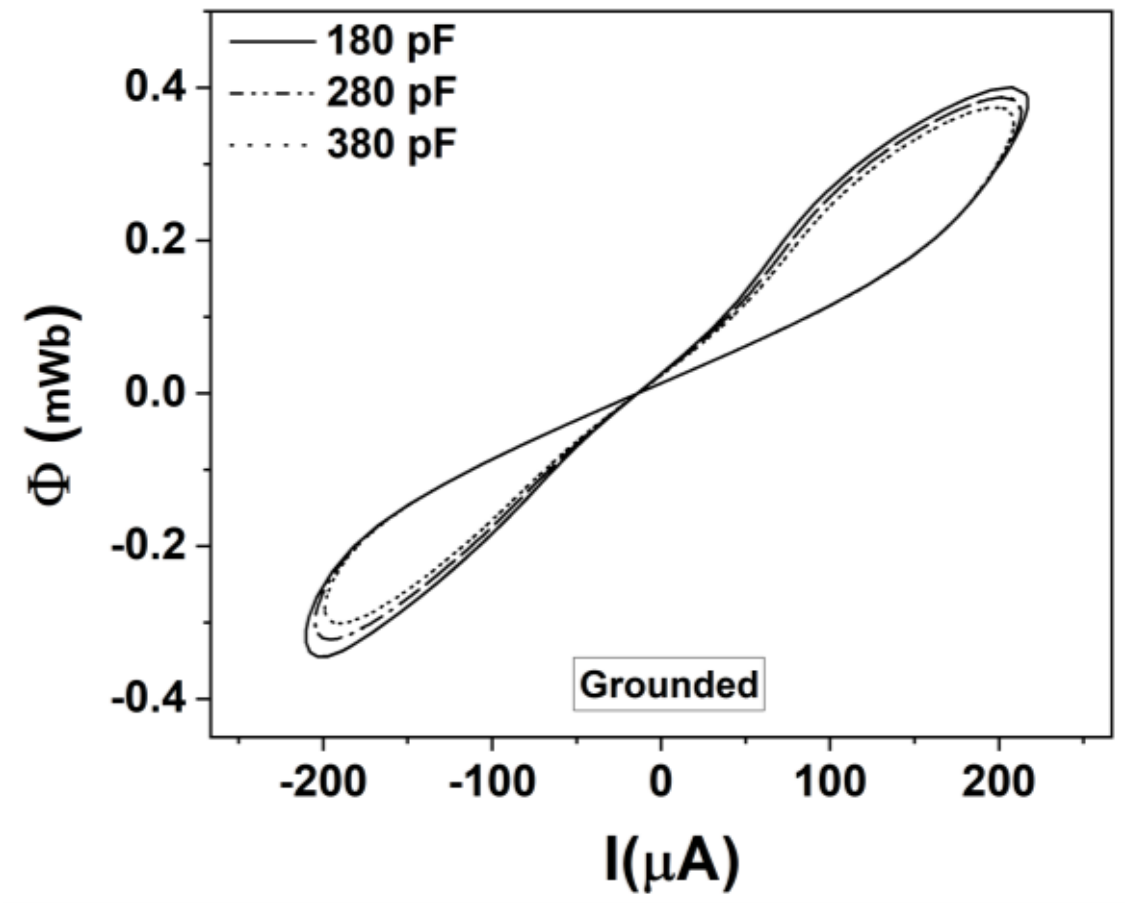

(c) Grounded

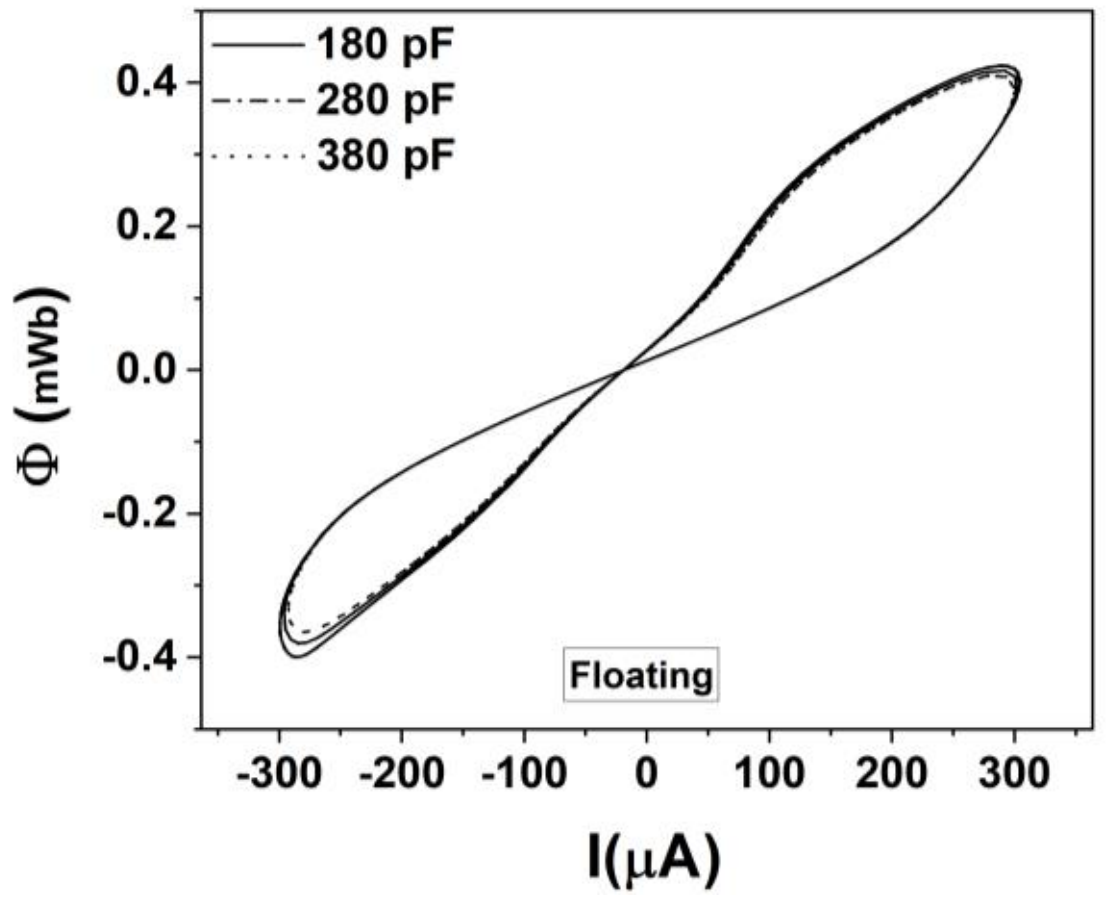

(d) Floating

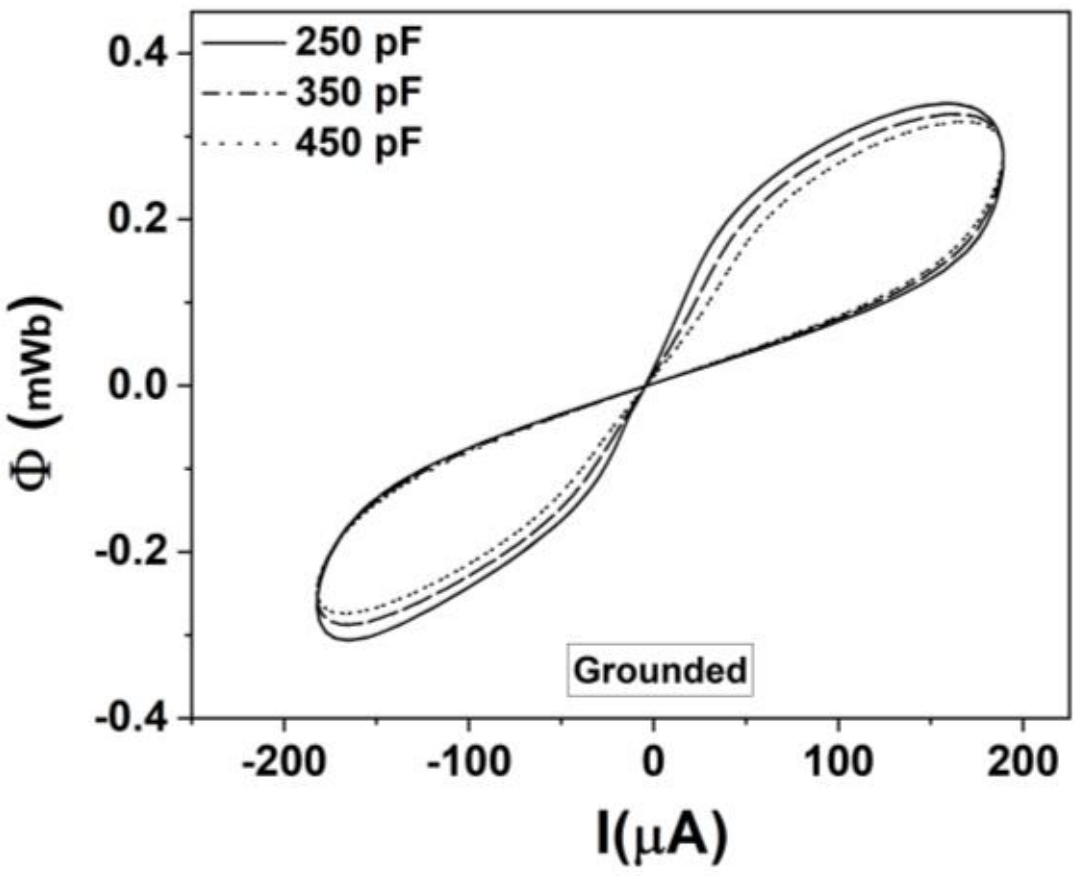

(e) Grounded

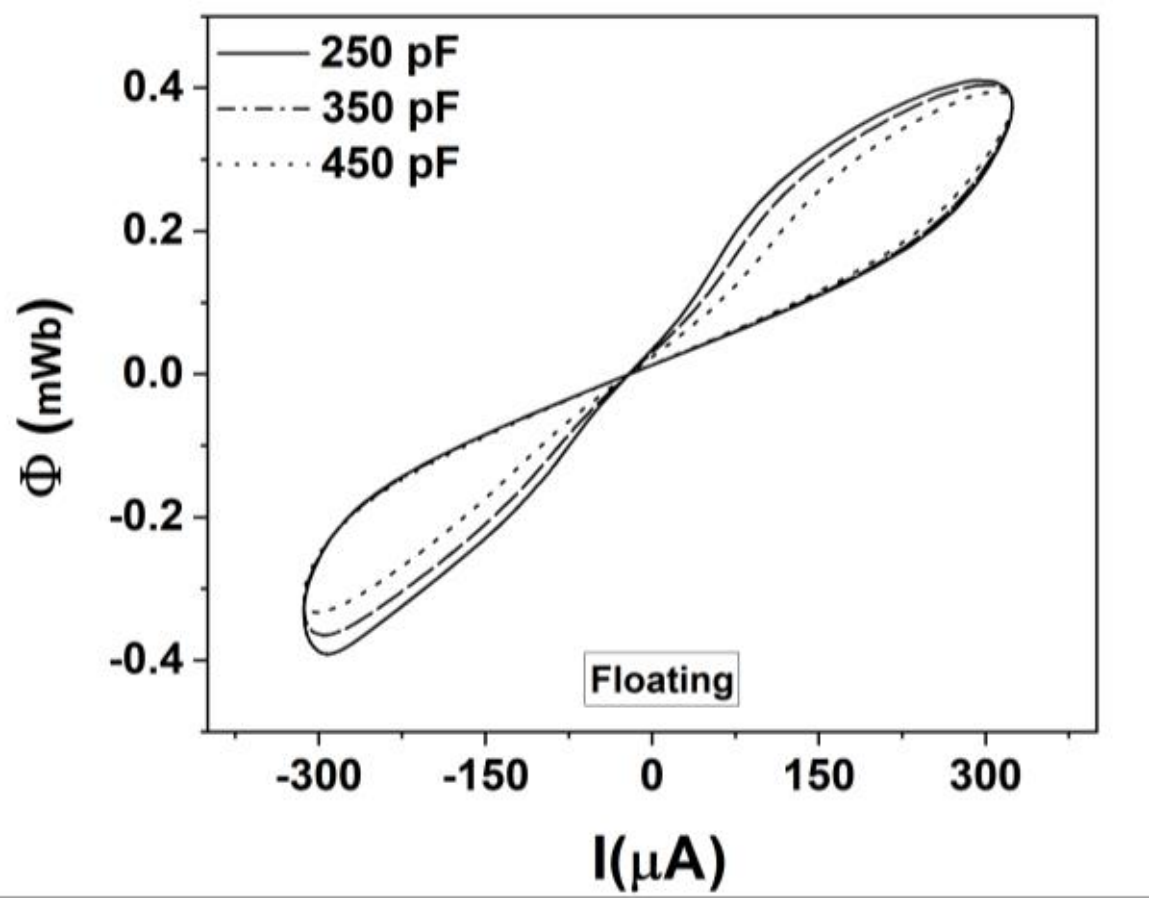

(f) Floating

**Figure. 11.** Φ-I characteristic curves for a sinusoidal current signal of $A_m$=140 mV and $V_{B4}$=500mV for (a) grounded meminductor for the variable frequency with $C_1$=180 pF, $C_2$=150pF, $I_b$=20 μA, (b) floating meminductor for the variable frequency at $C_1$=375 pF, $C_2$=150pF, $I_{b2}$=20 μA, (c) grounded meminductor for variable $C_2$ at 500 kHz (d) floating meminductor for variable $C_2$ at 800 kHz frequency, (e) grounded meminductor for variable $C_1$ at 500 kHz frequency, (f) floating meminductor for variable $C_1$ at 800 kHz frequency.

## 6. Layout, post-layout simulations, and Monte Carlo analysis

Layouts are obtained, and post-layout simulations are carried out for both grounded and floating topologies to check the effect of parasitics on the hysteresis. Further, the simulation results of the Monte Carlo analysis of proposed emulator circuits are also presented.

### *6.1. Pre and post-layout results for grounded and floating meminductor emulators*

Layouts of grounded and floating meminductor emulators are shown in Figure. 12. Layout areas occupied by grounded and floating meminductor emulators are 1845 $\mu m^2$ and 1874 $\mu m^2$, respectively. Figure. 13(a, b) shows the Φ-I characteristic curves for grounded meminductor emulators at 500 kHz and 1 MHz, respectively. Similarly, Figure. 13(c, d) shows the Φ-I characteristic curves for floating meminductor emulators at 100 kHz and 1 MHz operating frequencies, respectively. It is observed in Figure. 13 that the pre and post-layout results are in close agreement except for slight deviations due to parasitics present in active blocks.

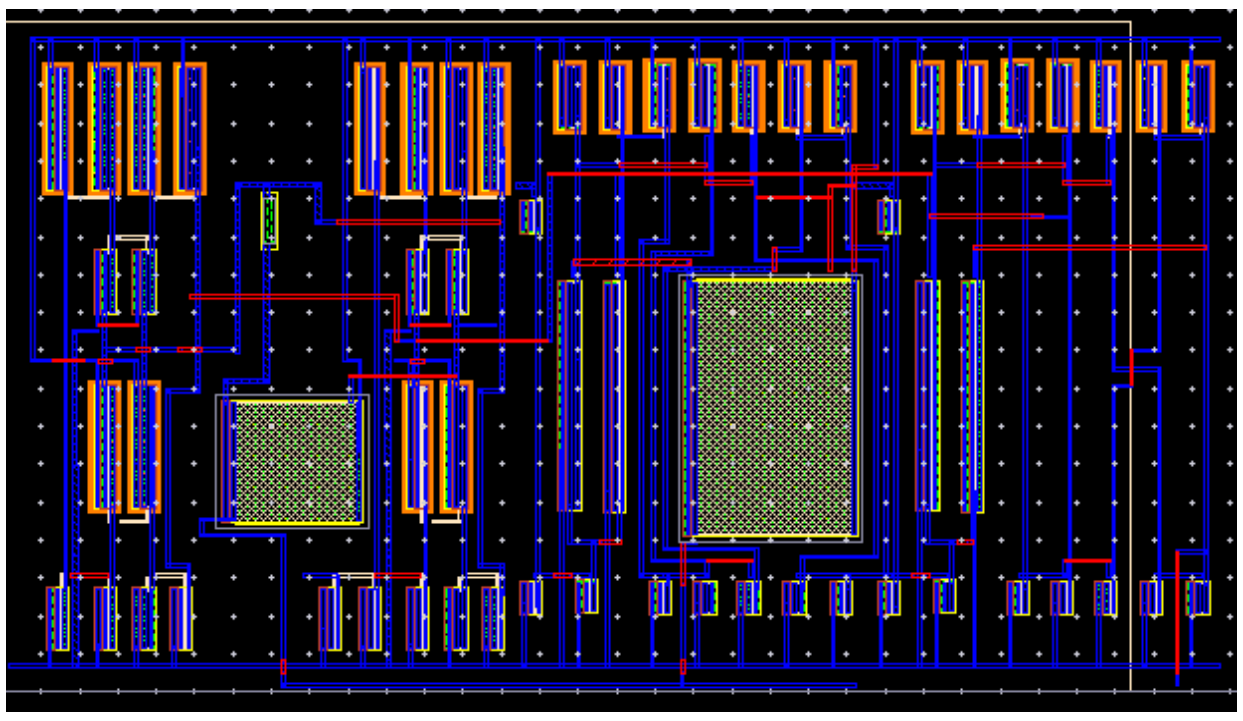

(a)

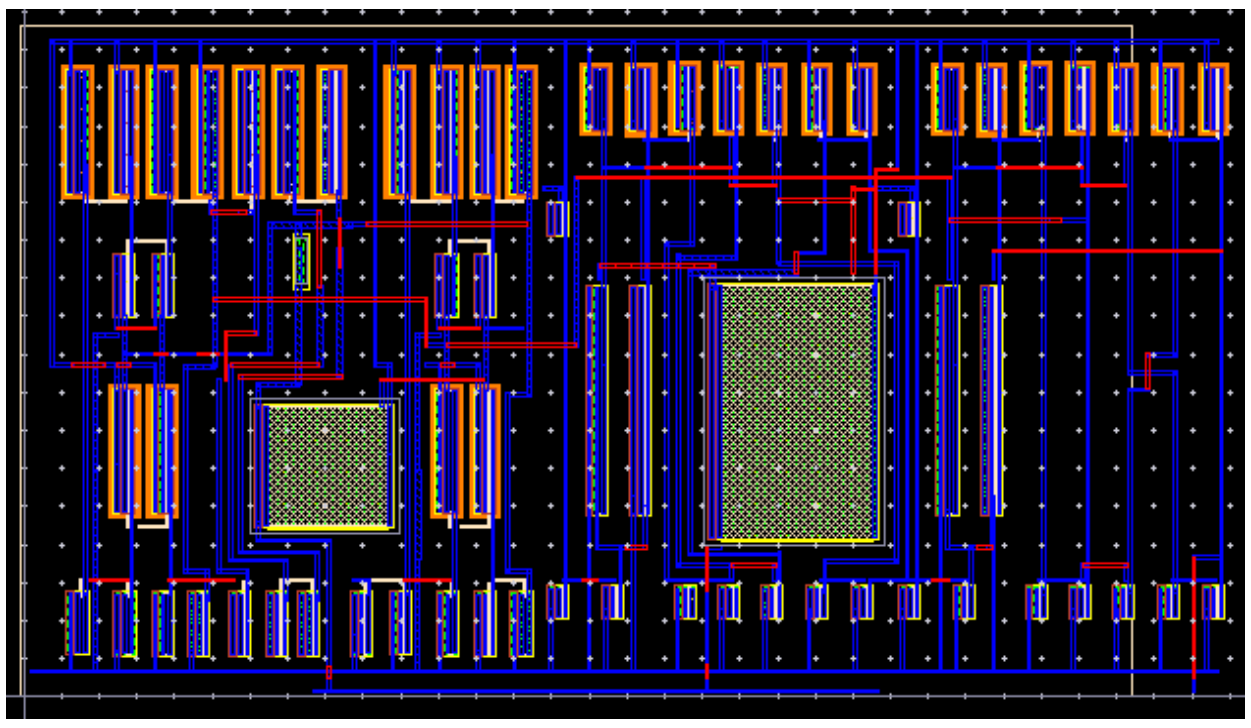

(b)

**Figure. 12.** Layout of meminductor emulators (a) grounded (b) floating.

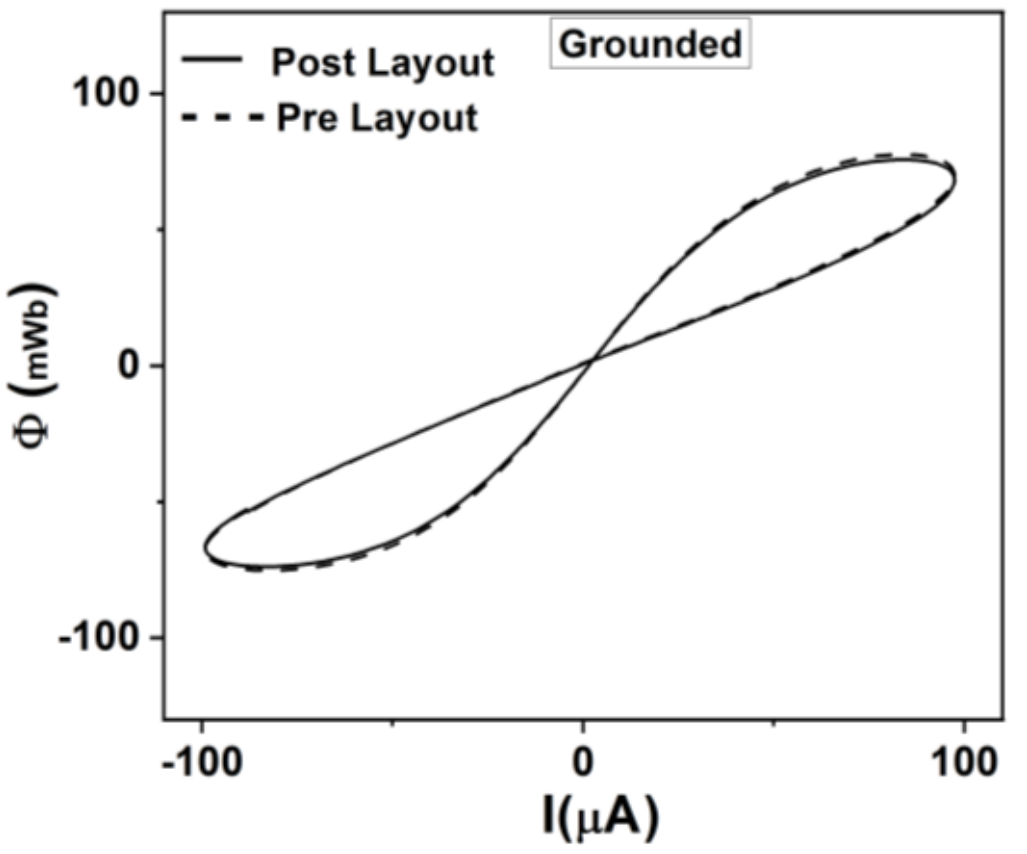

(a)

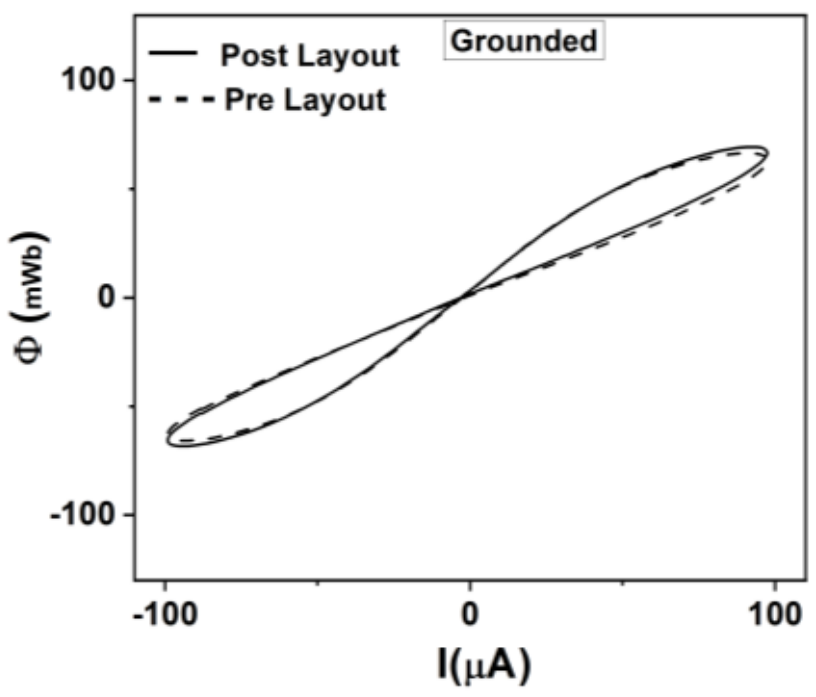

(b)

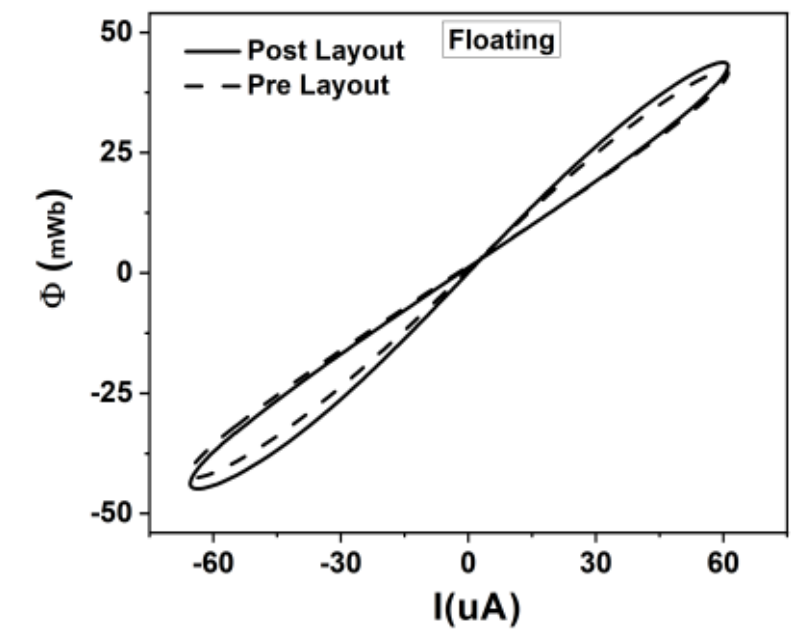

(c)

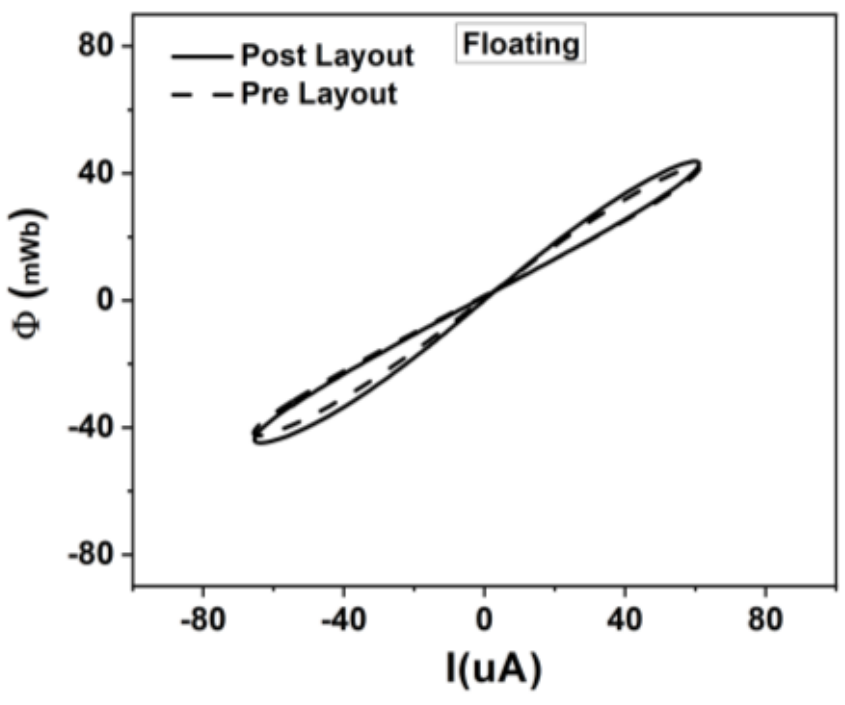

(d)

**Fig. 13.** Pre and Post layout simulation results of Φ-I characteristic plot for (a) grounded meminductor emulator at 500 kHz for $V_{B4}$ = 450 mV, $A_m$= 140 mV, $C_1$=48 pF, $C_2$= 80 pF and $I_b$=20 μA, (b) grounded meminductor emulator at 1 MHz for $V_{B4}$ = 450 mV, $A_m$= 140 mV, $C_1$= 8 pF, $C_2$=13 pF and $I_b$=20 μA, (c) floating meminductor emulator at 100 kHz for $V_{B4}$ = 500 mV, $A_m$= 200 mV, $C_1$=75 pF, $C_2$=150 pF and $I_{b2}$=18 μA, (d) floating meminductor emulator at 1 MHz for $V_{B4}$ = 500 mV, $A_m$= 200 mV, $C_1$=18 pF, $C_2$=28 pF, and $I_{b2}$=18 μA.

*6.2.Monte Carlo Analysis*

Post-layout Monte Carlo (MC) simulation for process mismatch at a frequency of 1 MHz for 200 simulation runs is performed for grounded topology. A similar MC analysis is performed for floating topology at a frequency of 200 kHz for 200 simulation runs. MC results for the hysteresis loop in both grounded and floating topology are shown in Fig 14. It reveals that the proposed circuit is affected by process mismatch; however, the hysteresis loop closely retains the original form.

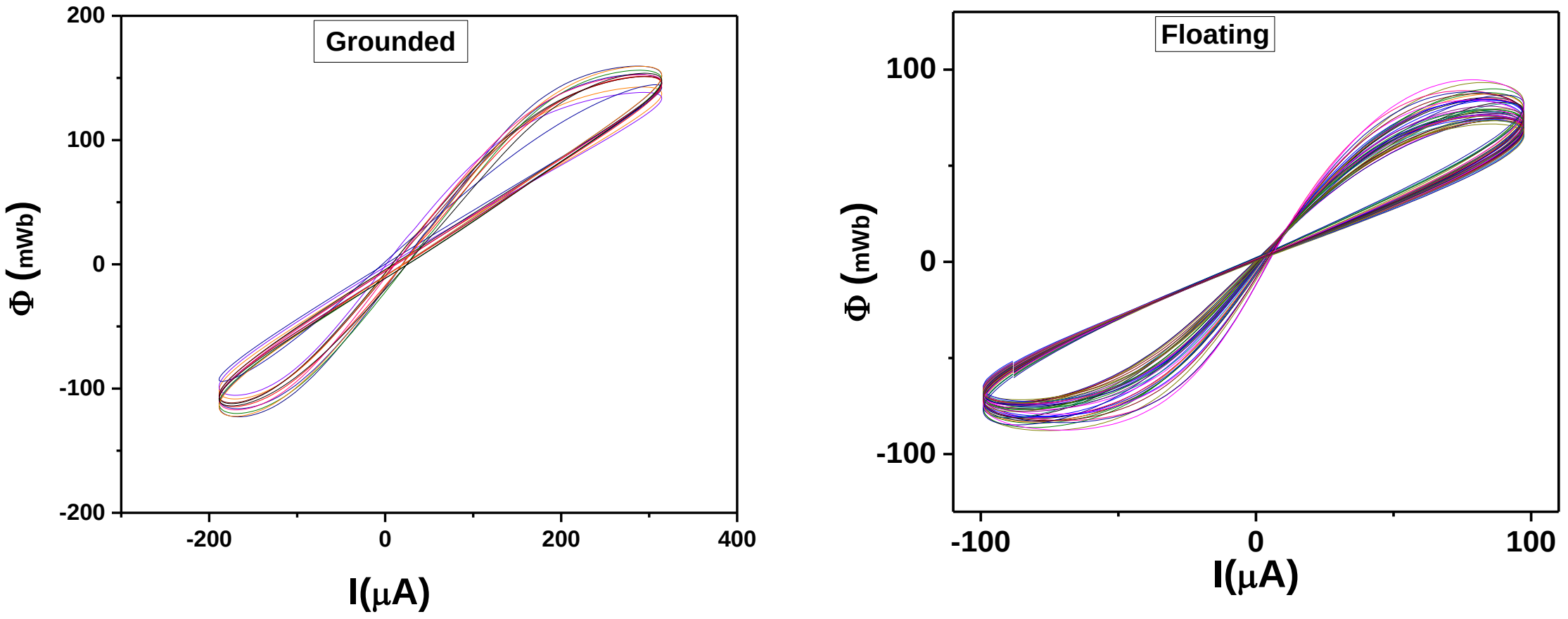

(a) (b)

**Fig. 14.** Hysteresis loop of MC result for 200 simulation runs for (a) grounded topology and (b) floating topology.

## 7. Nonideality Analysis

Non-ideal transfer gains and parasitics of active building blocks will have an impact practically. Sections 6.1 discusses the effect due to non-ideal transfer gains, and sections 6.2 and 6.3 discuss the effect due to OTA and CCII/CCCII parasitics.

*7.1 Nonideality effect of OTA transconductance gain and CCII/CCCII current transfer gains*

Due to the OTA's non-ideal transfer gain, the port relationship is modified as follows:

$$I_{0\pm} = \pm\gamma G_m(V_{in+} - V_{in-}) = \pm\gamma G_m V_{in} \tag{35}$$

$\gamma$ is the non-ideal transconductance gain coefficient from the input terminal to the output terminal of OTA, which is ideally considered unity.

Similarly, the port relationships of CCII and CCCII due to non-ideal transfer gains modify as follows:

$$\begin{bmatrix} V_X \\ I_Z \\ I_Y \end{bmatrix} = \begin{bmatrix} \beta_1 & 0 & 0 \\ 0 & \alpha_1 & 0 \\ 0 & 0 & 0 \end{bmatrix} \begin{bmatrix} V_Y \\ I_X \\ V_Z \end{bmatrix} \quad \text{and} \tag{36}$$

$$\begin{bmatrix} V_X \\ I_Z \\ I_Y \end{bmatrix} = \begin{bmatrix} \beta^{(j)} & R_X & 0 \\ 0 & \alpha^{(j)} & 0 \\ 0 & 0 & 0 \end{bmatrix} \begin{bmatrix} V_Y \\ I_X \\ V_Z \end{bmatrix}, \tag{37}$$

Where j=1 stands for the first CCCII and j=2 for the second CCCII. $\alpha$ and $\beta$ are current and voltage transfer gains from X to Z and Y to X terminals for CCII/CCCII, respectively. Ideally, $\alpha$ and $\beta$ are unity.

The routine analysis of Fig. 4 considering non-ideal port relationships from (35), (36), and (37) for grounded meminductor configuration results in the following:

$$\frac{\phi_{in}}{I_{in}} = \frac{R_1 C_2 \beta^{(2)} \alpha_1 \alpha^{(2)}}{\frac{k\gamma_1}{\sqrt{2}}(V_{ss}+2V_{th}) \mp \frac{k\gamma_1\gamma_2}{\sqrt{2}}\left(\frac{G_{m4}(1+sC_2R_{X2})\phi_{in}}{\beta^{(2)}\alpha_1\alpha^{(2)}sC_1R_1C_2}\right)} \frac{1}{1+sC_2R_{X2}} \tag{38}$$

Similar analysis for the floating topology of Fig. 5 results in :

$$\frac{\alpha^{(1)}\beta^{(2)}V_{in1}-V_{in2}}{I_{in}} = \frac{sR_{X2}C_2}{\frac{k\gamma_1\alpha^{(2)}}{\sqrt{2}}(V_{ss}+2V_{th}) \mp \frac{k\alpha^{(2)}\gamma_1\gamma_2}{\sqrt{2}}\left(\frac{G_{m4}\alpha^{(2)}\int(\alpha^{(1)}\beta^{(2)}V_{in1}-V_{in2})dt}{sC_1R_{X2}C_2}\right)} \tag{39}$$

[where $V_{in1} - V_{in2} = V_{in}$ and $\int (V_{in1} - V_{in2})dt = \phi_{in}$]

$\gamma_1$ *and* $\gamma_2$ are the transconductance gains of grounded and floating meminductors' first and second OTAs. It is observed from (38) and (39) that non-ideal transfer gains affect meminductance.

*7.2 Nonideality effect due to device parasitics on grounded meminductor emulator*

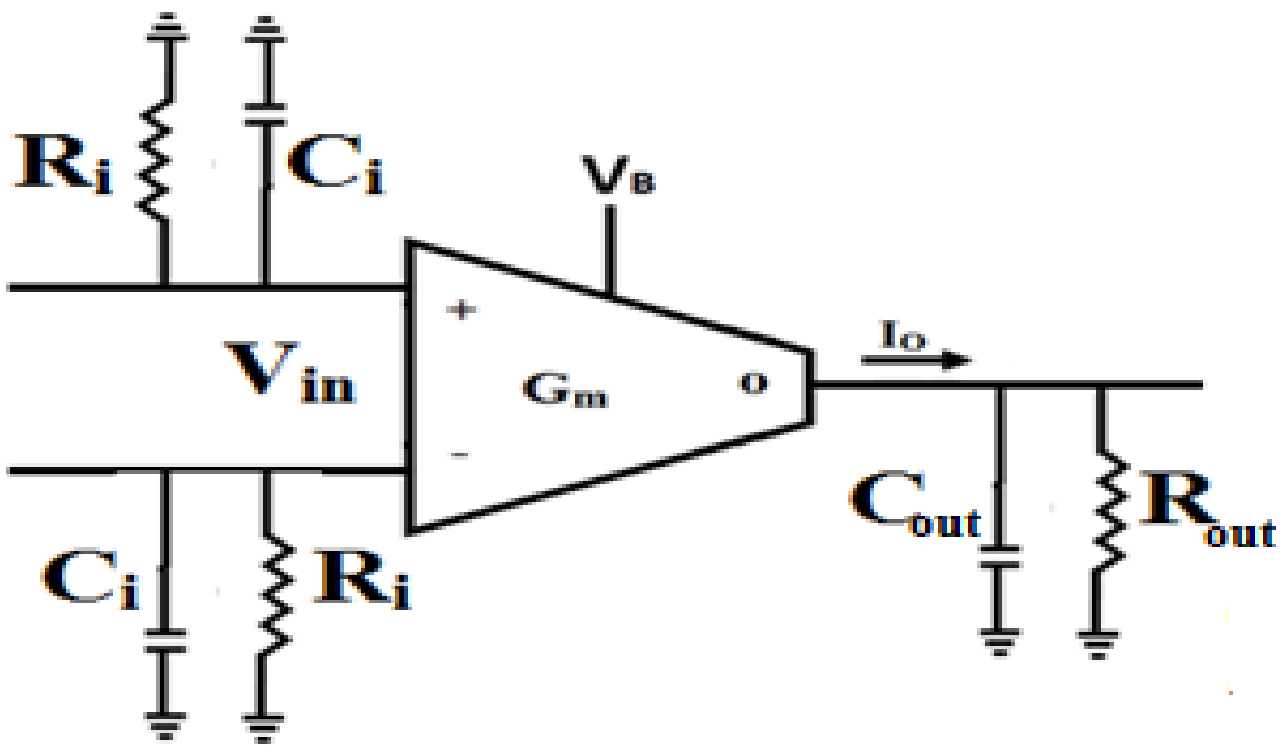

**Fig. 15.** Non-ideal model of OTA [25-26] .

Fig. 15 shows the non-ideal model of OTA where ($R_i$, $C_i$) and ($R_0$, $C_0$) are input and output parasitic capacitances and resistances, respectively. The capacitances across the input ports are assumed to be equal.

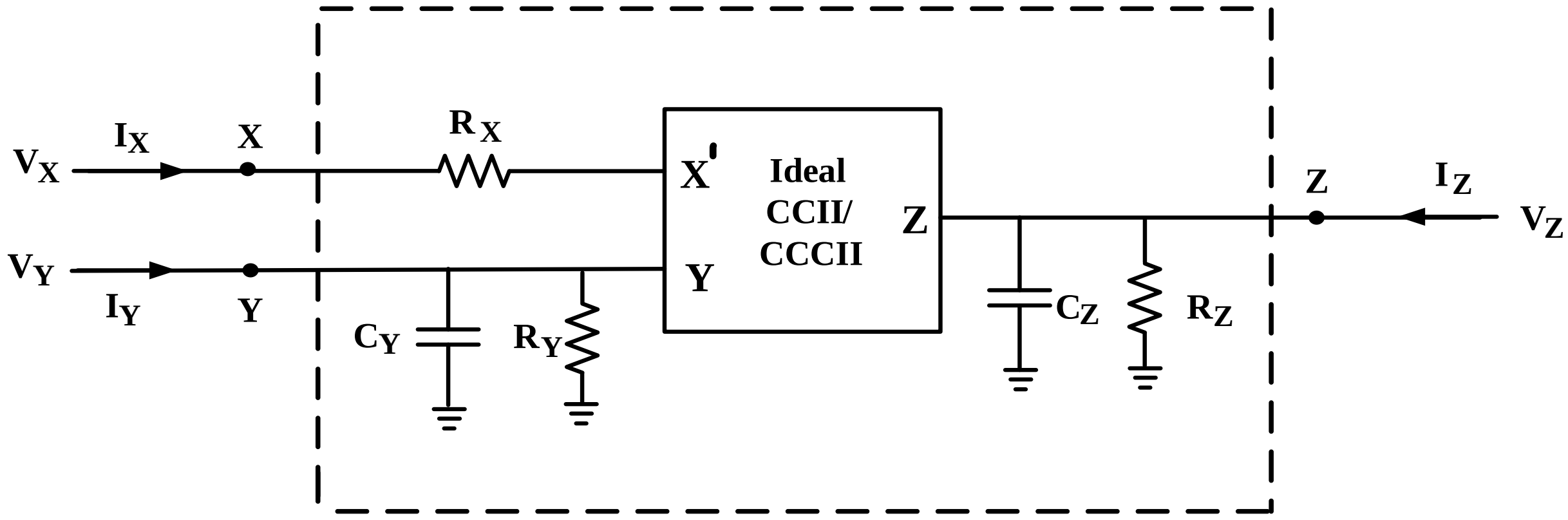

**Fig. 16.** Non-Ideal model of the CCII/CCCII [38].

Fig. 16 shows the non-ideal model of CCII/CCCII. $R_X$ represents a low-value parasitic resistance at the X terminal for CCII., whereas $R_X$ for CCCII is a variable intrinsic resistance. At terminal Y and Z, the parasitic components are in parallel (i.e., $R_Y \parallel C_Y$ and $R_Z \parallel C_Z$), where $R_Y$ and $R_Z$ are of high value, and $C_Y$ and $C_Z$ are of low value.

Fig. 17 shows the non-ideal model with device parasitics of proposed grounded meminductor emulators of Fig. 4. Let the impedances be:

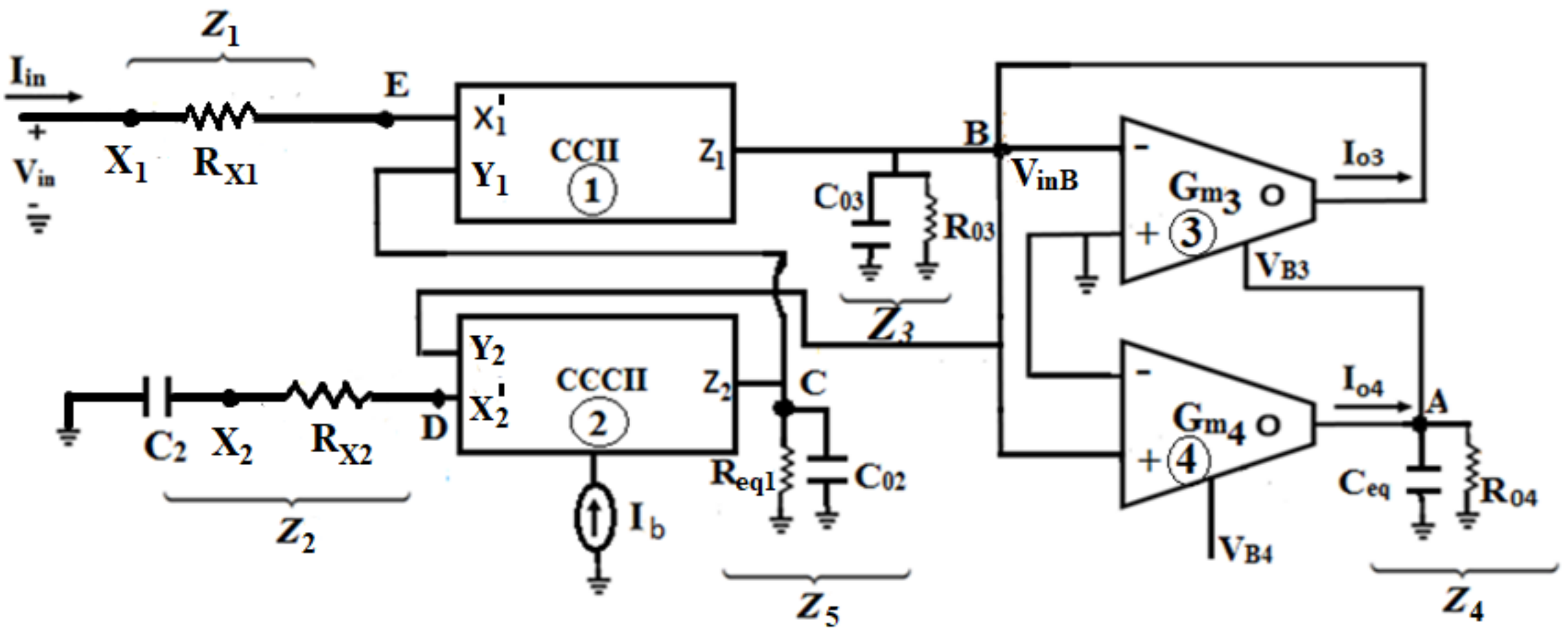

**Fig. 17.** Non-ideal model of grounded meminductor emulator.

$$Z_1 = (R_{X1}), \quad Z_2 = (R_{X2} + X_{C2}), \quad Z_3 = (R_{03} \parallel X_{C03}),$$

$$Z_4 = (R_{04} \parallel X_{Ceq}) \text{ and } Z_5 = (R_{eq1} \parallel X_{C02}), \tag{40}$$

where

$$C_{eq} = (C_1 + C_{out4}), \; R_{03} = (R_{i3} \parallel R_{i4} \parallel R_{Z1} \parallel R_{Y2} \parallel R_{out3}), R_{eq1} = (R_1 \parallel R_{Y1} \parallel R_{z2}),$$

$$C_{03} = (C_{i3} \parallel C_{i4} \parallel C_{Z1} \parallel C_{Y2} \parallel C_{out3}), C_{02} = (C_{Z2} \parallel C_{Y1}) \text{ and } R_{o4} = R_{out4} \tag{41}$$

Where $R_{i3}$, $C_{i3}$, and $R_{i4}$, $C_{i4}$ are parasitic resistances and capacitances at the inputs of the first OTA ($G_{m3}$) and second OTA ($G_{m4}$), respectively. Similarly, $R_{out3}$, $C_{out3}$, and $R_{out4}$, $C_{out4}$ are parasitic resistances and capacitances at the first OTA ($G_{m3}$) and second OTA ($G_{m4}$) output, respectively.

$$V_{Y2} = V_{inB} = V_{X2} - I_{X2}R_{X2} = -I_{X2}X_{C2} - I_{X2}R_{X2} = -I_{X2}Z_2 \quad \text{(by port relationship)} \tag{42}$$

Also, on applying KCL at node C and by port relationship

$$I_{X2} = I_{Z2} = \frac{-V_{Y1}}{Z_5} = \frac{-(V_{X1})}{Z_5} = \frac{-(V_{in})}{Z_5} \tag{43}$$

Using (43) in (42) results in

$$V_{in} = \psi V_{inB}, \quad \text{where } \psi = \left(\frac{Z_5}{Z_2}\right) \tag{44}$$

On applying KCL at node B

$$I_{in} = I_{Z1} = I_{O3} - \frac{V_{inB}}{Z_3} \tag{45}$$

Using (44) and (45) results in

$$\frac{I_{in}}{V_{in}} = \left(\frac{1}{\psi}\right)\left(\frac{I_{O3}-\frac{V_B}{Z_3}}{V_B}\right) = \left(\frac{1}{\psi}\right)\left(\frac{I_{O3}}{V_B}\right)\left(1-\frac{V_{inB}}{I_{O3}Z_3}\right) \tag{46}$$

Further analysis using (1) and $G_{m3} = \frac{-I_{O3}}{V_{inB}}$ results in the following;

$$\frac{I_{O3}}{V_{inB}} = \frac{k}{\sqrt{2}}(V_{ss}+2V_{TH}) - \frac{kZ_4G_{m4}V_{in}}{\sqrt{2}\,\psi} \tag{47}$$

So the inverse meminductance equation after substituting (47) in (46) becomes for both the incremental and decremental topologies as

$$\frac{I_{in}}{V_{in}} = \frac{1}{\psi}\left(\frac{k}{\sqrt{2}}(V_{ss}+2V_{th}) \mp \frac{kZ_4G_{m4}V_{in}}{\sqrt{2}\,\psi}\right)\left(1-\frac{V_B}{I_{O3}Z_3}\right) \tag{48}$$

Typical practical values of CMOS OTA parasitic obtained from routine analysis and [38] can be assumed approximately as, $R_{i3} = R_{i4} = \infty$, $R_o = 1M\Omega, C_i = 50fF$, $C_0 = 100fF$. Similarly, values of CCII/CCCII obtained from routine analysis and [40] can be assumed to be a few ohms for Rx, a few hundreds of MΩ for $R_Y$, and few MΩ for $R_Z$, and the Cy, Cz are in the range of a few femtofarads. If the operating frequency is within the range of a few MHz, then we may use different terms as follows: $Z_2 \approx R_{X2} + \frac{1}{sC_2}$, $Z_4 \approx 1/sC_1$, $Z_5 \approx R_1$=few Ω, $\psi = \frac{sR_1C_2}{1+sC_2R_{X2}}$, and $Z_3 \approx \infty$, and $\left[\frac{V_{inB}}{I_{O3}Z_3}\right] \approx 0$ as $(Z_3 \approx \infty)$. The substitution of these values in (48) results in the following:

$$\frac{V_{in}}{I_{in}} = \frac{sR_1C_2}{\frac{k}{\sqrt{2}}(V_{ss}+2V_{th}) \mp \frac{k}{\sqrt{2}}\left(\frac{G_{m4}(1+sC_2R_{X2})\phi_{in}}{sC_1R_1C_2}\right)}\frac{1}{1+sC_2R_{X2}}$$

(49)This is similar to (14). Hence it can be concluded that the effect of parasitic on the proposed grounded circuit in the frequency range of a few MHz is negligible.

### *6.3 Nonideality effect due to device parasitic on floating meminductor emulator*

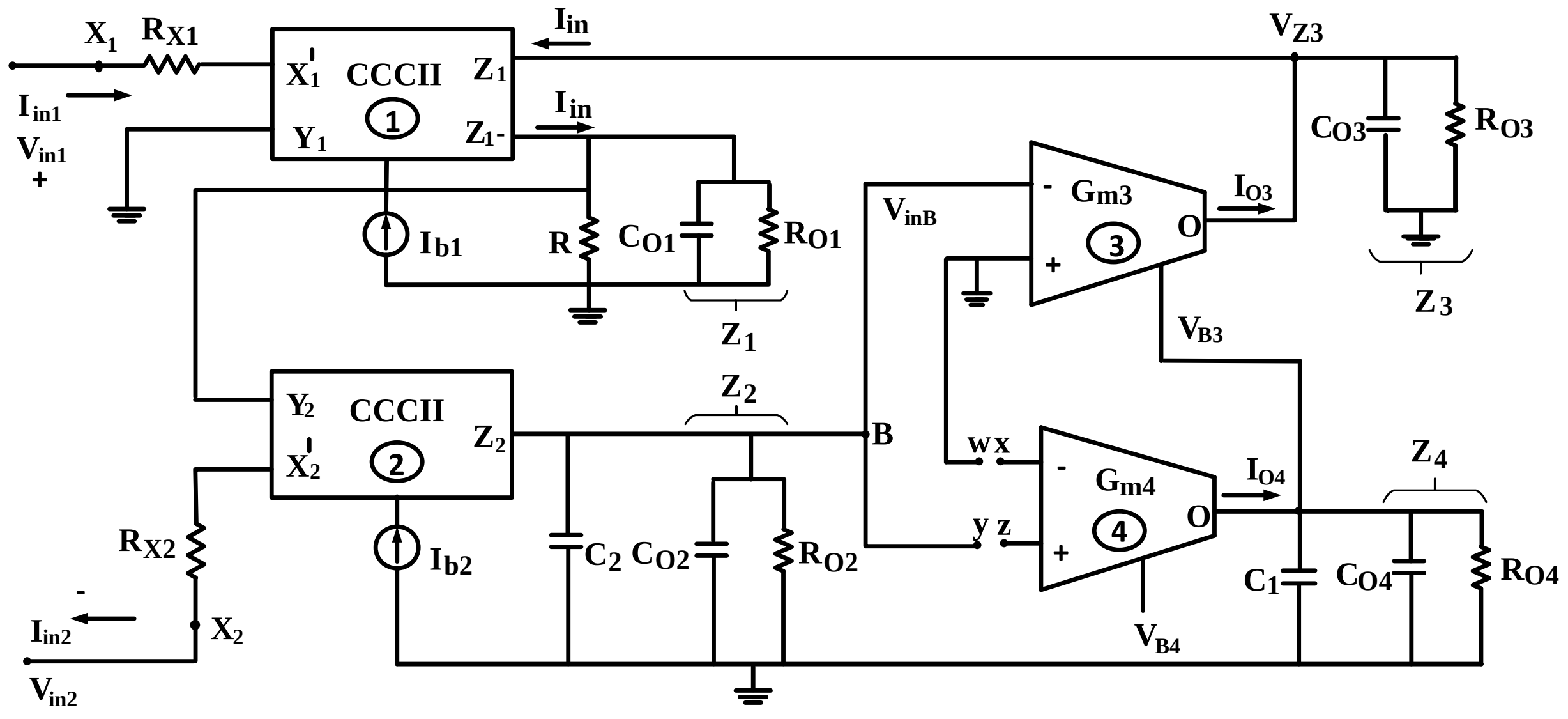

**Fig. 18.** Non-Ideal model of proposed floating meminductor emulator.

In Fig. 18, let the equivalent impedances at nodes are as follows,

$$Z_1 = (R_{01} \parallel X_{C01}) \parallel R, \quad Z_2 = (R_{02} \parallel X_{C02}) \parallel X_{C2} ,$$

$$Z_3 = (R_{03} \parallel X_{C03}) \quad \text{and} \quad Z_4 = (R_{04} \parallel X_{C04}) \parallel X_{C1} \tag{50}$$

and

$$C_{01} = (C_{Z1-} + C_{Y2}), \quad C_{02} = (C_{Z2} + C_{i3} + C_{i4}), C_{03} = (C_{out3} + C_{Z1}), C_{04} = (C_{out4})$$

$$R_{01} = (R_{Z1-} \parallel R_{Y2}), \quad R_{02} = (R_{Z2} \parallel R_{i3} \parallel R_{i4}), \; R_{03} = (R_{Z1} \parallel R_{out3}), \mathrm{R}_{04} = (R_{out4}) \tag{51}$$

Also, $$V_{Y2} = V_{z1-} = \frac{V_{in1}}{R_{X1}} Z_1 \text{ (by port relationship)} \tag{52}$$

By applying the port relationship at the $X_2$ terminal of CCCII and using (52), we get:

$$V_{X2} = V_{in2} = V_{Y2} + I_{X2}R_{X2} = \frac{V_{in1}}{R_{X1}} Z_1 + I_{X2}R_{X2} \tag{53}$$

Moreover at node B, $-I_{Z2}\, Z_2 = V_{inB}$, or, $I_{Z2} = I_{X2} = \frac{-V_{inB}}{Z_2}$ (54)

Thus, from (53) and (54), $\frac{V_{in1}}{R_{X1}} Z_1 - V_{in2} = V_{inB} \frac{R_{X2}}{Z_2}$ (55)

Further as ; $I_{in} = I_{Z1} = I_{O3} - \frac{V_{Z3}}{Z_3}$ (56)

Hence, from (55) and (56)

$$\frac{\frac{V_{in1}}{R_{X1}}Z_1 - V_{in2}}{I_{in}} = \frac{V_{inB}}{(I_{O3} - \frac{V_{Z3}}{Z_3})}\frac{R_{X2}}{Z_2} \quad (57)$$

Or $$\frac{\frac{V_{in1}}{R_{X1}}Z_1 - V_{in2}}{I_{in}} = \frac{V_{inB}}{I_{O3}\left(1 - \frac{V_{Z3}}{I_{O3}Z_3}\right)}\frac{R_{X2}}{Z_2} \quad (58)$$

Further, As $V_{B3} = Z_4 I_{O4} = Z_4 G_{m4} V_{inB}$ (59)

Hence from (55), by putting the value of $V_{inB}$ in (59), $V_{B3}$ becomes ;

$$V_{B3} = (G_{m4}Z_4)\left(\frac{Z_2}{R_{X2}}\right)\left(\frac{V_{in1}}{R_{X1}}Z_1 - V_{in2}\right) \quad (60)$$

Also, as, $G_{m3} = \frac{-I_{O3}}{V_{inB}}$ (61)

Further from (1) and (60),

$$G_{m3} = \frac{k}{\sqrt{2}}(V_{B3} - V_{ss} - 2V_{th}) = \frac{k}{\sqrt{2}}\left(\{(G_{m4}Z_4)\left(\frac{Z_2}{R_{X2}}\right)\left(\frac{V_{in1}}{R_{X1}}Z_1 - V_{in2}\right)\} - V_{ss} - 2V_{th}\right)$$

Or from (61);

$$\frac{I_{O3}}{V_{inB}} = -\frac{k}{\sqrt{2}}\left(\{(G_{m4}Z_4)\left(\frac{Z_2}{R_{X2}}\right)(\frac{V_{in1}}{R_{X1}}Z_1 - V_{in2})\} - V_{ss} - 2V_{th}\right)$$

$$\frac{I_{O3}}{V_{inB}} = \frac{k}{\sqrt{2}}\left(V_{ss} + 2V_{th} - \{(G_{m4}Z_4)\left(\frac{Z_2}{R_{X2}}\right)\left(\frac{V_{in1}}{R_{X1}}Z_1 - V_{in2}\right)\}\right) \quad (62)$$

On substituting (62) in (58) results for both the incremental and decremental topologies as;

$$\frac{\frac{V_{in1}}{R_{X1}}Z_1 - V_{in2}}{I_{in}} = \frac{\frac{R_{X2}}{Z_2}}{\frac{k}{\sqrt{2}}\left(V_{ss} + 2V_{th} \mp (G_{m4}Z_4)\left(\frac{Z_2}{R_{X2}}\right)(\frac{V_{in1}}{R_{X1}}Z_1 - V_{in2})\right)}\left(1 - \frac{V_{Z3}}{I_{O3}Z_3}\right) \quad (63)$$

It is inferred from (63) that meminductance will be affected by parasites, however on applying the typical practical values of CMOS OTA parasitic obtained from routine analysis and [37] can be assumed approximately as; $R_i = \infty, R_o = 1M\Omega, C_i = 50fF, C_0 = 100fF$, while that of CCII/CCCII as discusses in above. If the operating frequency is within the range of a few MHz (say 10 MHz), then we may approximate the following as follows: $Z_2 = X_{C2}$, $Z_4 = X_{C1}$, $Z_1 = R$, $R$=$R_{X1}$, $Z_3 \approx$ very high, and hence, $\left[\frac{V_{Z3}}{I_{O3}Z_3}\right] \approx 0$. The substitution of these terms in (63 for both incremental and decremental topologies results in:

$$\frac{V_{in}}{I_{in}} = \frac{sR_{X2}C_2}{\frac{k}{\sqrt{2}}(V_{SS}+2V_{th})\mp\frac{k}{\sqrt{2}}\left(\frac{G_{m4}\phi_{in}}{sC_1R_{X2}C_2}\right)} \quad (64)$$

This is similar to (32) and (33). Hence it can be concluded that the effect of parasitic on the proposed floating meminductor in the frequency range of low to 10 MHz is negligible.

## 8. Application of meminductor as amplitude modulator (AM)

An AM modulation scheme with a meminductor is carried out as an application of the proposed meminductive device. Schematic of various circuits along with a floating meminductor emulator is shown in Figure. 19. In Figure. 19(a), a multifunction filter [39] using OTA is given, which can implement both bandpass filter (BPF) and low pass filter (LPF) responses using the components, as given in Table 4, for $Y_1$, $Y_2$, $Y_3$, $Y_4$, and $Y_5$. Current mode (CM) BPF and LPF filters are used for modulation and demodulation, respectively. The meminductance of the meminductor shown in Figure. 19(b) is controlled by the low-frequency message signal $V_m(t)$, due to which message gets imposed on the high-frequency carrier signal $V_c(t)$. The output is filtered out by the bandpass filter in Figure. 19(a) centered at the carrier frequency to obtain an amplitude-modulated wave. The circuit in Figure. 19(c) is used to demodulate the AM signal to recover the message signal.

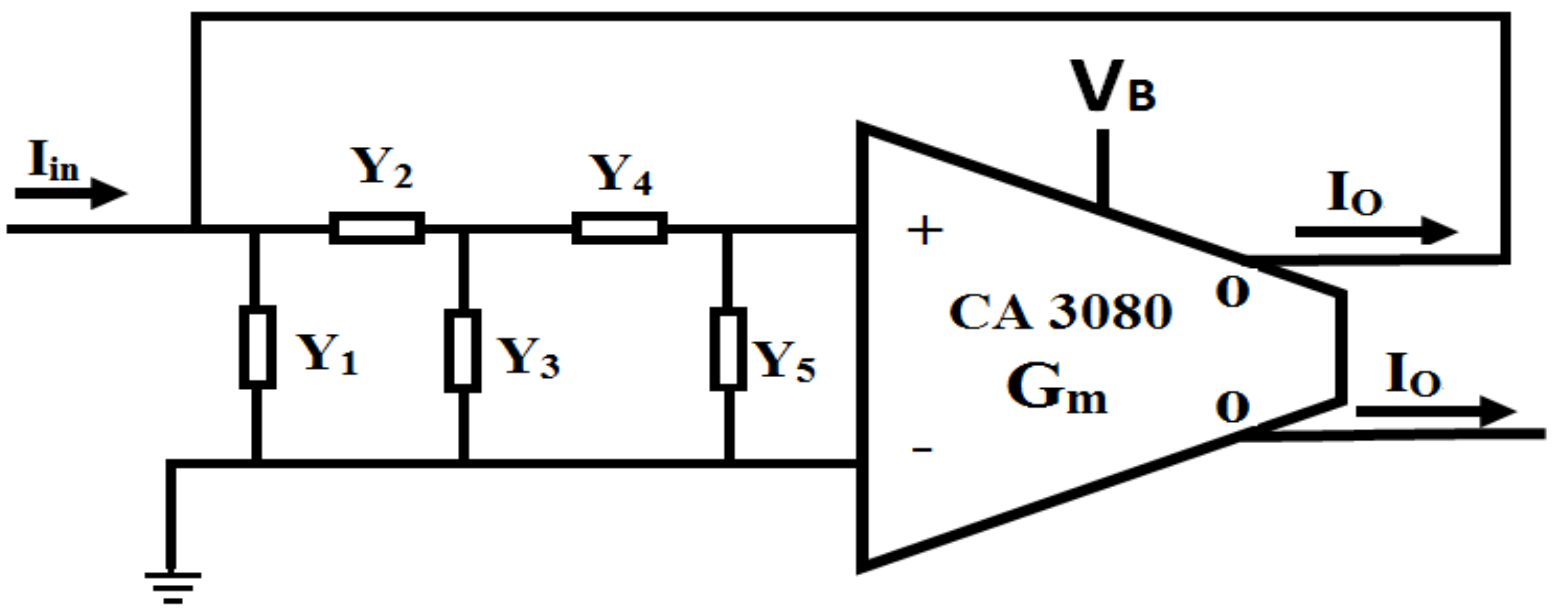

(a)

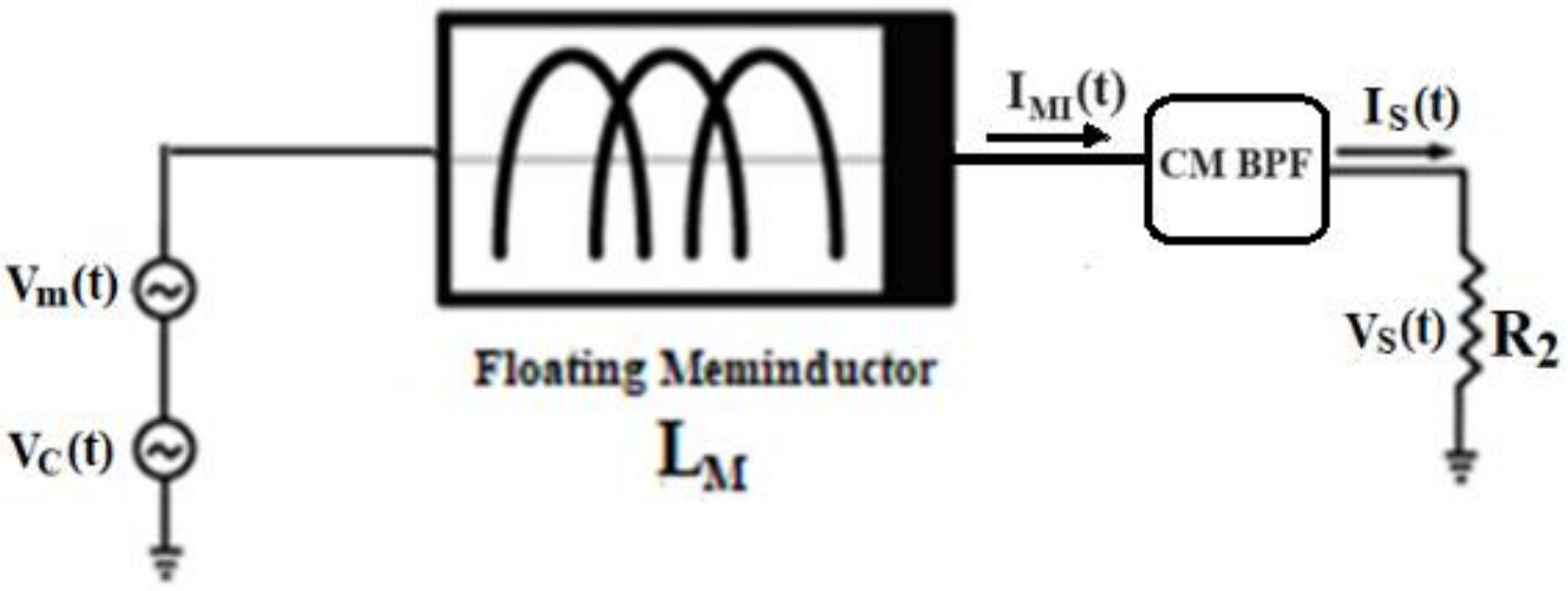

(b)

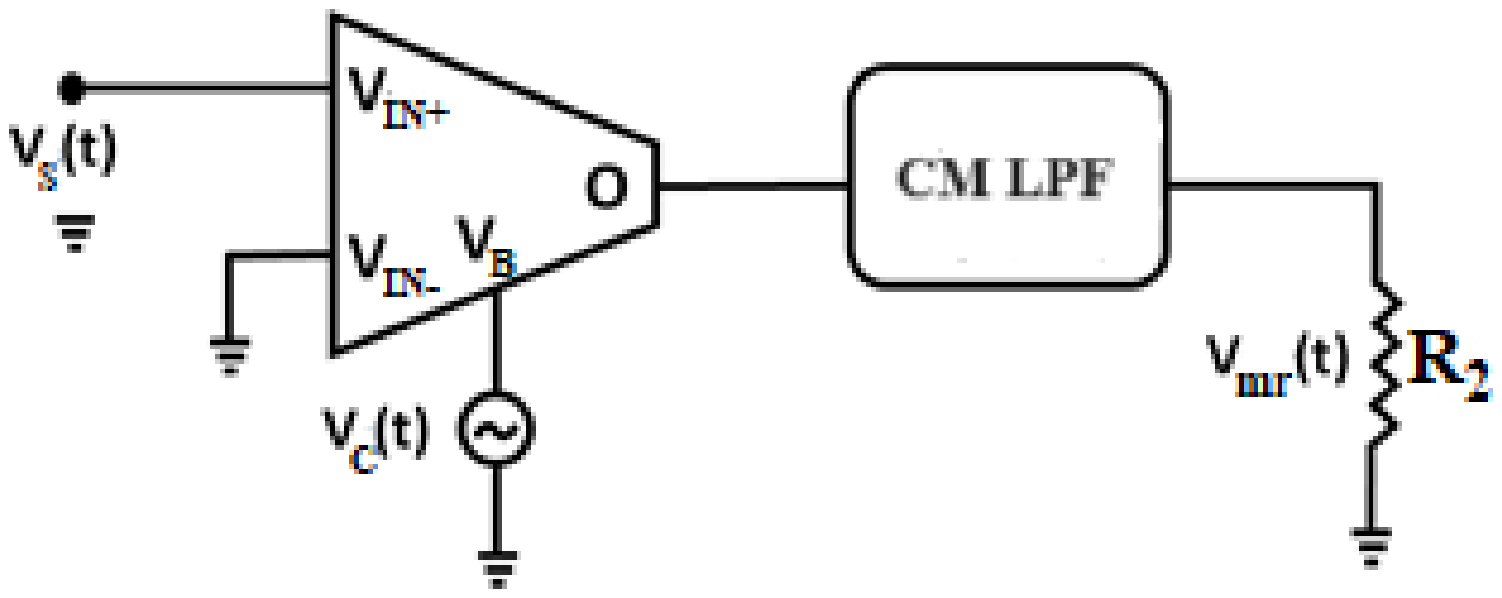

(c)

**Figure. 19.** Block diagram of (a) current mode multifunction filter, (b) AM modulator circuit using floating meminductor emulator and filter, (c) coherent demodulator circuit using OTA.

**Table 4.** Specification for multimode filter components

| | **$Y_1$** | **$Y_2$** | **$Y_3$** | **$Y_4$** | **$Y_5$** |
|---|---|---|---|---|---|
| Low pass | C1 | R2 | C3 | ∞ | 0 |
| Band pass | R1 | R2 | C3 | C4 | R5 |

*8.1 Analysis of amplitude modulator and demodulator*

$V_C(t)$ are given by

$$V_m(t) = A_m \cos(\omega_m t);\;\; V_C(t) = A_C \cos(\omega_c t) \quad (65)$$

The flux imposed on the meminductor is given by

$$\varphi_{in} = \int V_{in}(t)dt = \int (V_m(t) + V_C(t))dt = \frac{A_m \sin(\omega_m t)}{\omega_m} + \frac{A_C \sin(\omega_C t)}{\omega_C} \quad (66)$$

Also, $V_{in}(t) = A_m \cos(\omega_m t) + A_C \cos(\omega_c t)$ (67)

Thus, the output current of the meminductor, $I_{MI}(t)$, of Figure 19(b) results as:

$$I_{MI}(t) = V_B \frac{k}{\sqrt{2}}\left(\frac{G_{m4}}{C_1}\int V_B(t)dt - V_{ss} - 2V_{th}\right) \quad (68)$$

Now using (26), (66), and (67) in (68) and passing it through BPF yields;

$$I_s(t) = M_1 sin\omega_c t + M_2[sin(\omega_c + \omega_m) + sin(\omega_c - \omega_m)] + M_3[sin(\omega_c + \omega_m) - sin(\omega_c - \omega_m)] \quad (69)$$

Where,

$$M_1 = -\frac{k(V_{SS}+2V_{th})A_C}{\sqrt{2}R_{X2}C_2\omega_C},\ M_2 = \frac{A_C A_m G_{m4} k}{2\sqrt{2}C_1\omega_C}\left(\frac{1}{sR_{X2}C_2}\right)^2 \text{ and } M_3 = \frac{A_C A_m G_{m4} k}{2\sqrt{2}C_1\omega_m}\left(\frac{1}{sR_{X2}C_2}\right)^2$$

It is evident that (69) is in the form of components of the upper sideband, lower sideband, and carrier of standard AM expression. In order to recover the message signal from the modulated signal, a coherent product demodulator is used. The demodulator circuit is realized with OTA as a multiplier cascaded with an OTA-based low pass filter, as shown in Figure. 19(c).

The voltage message signal $V_m(t)$ and carrier signal $V_C(t)$ are taken, respectively as

$$V_m(t) = A_m\, cos(\omega_m t); \qquad V_C(t) = A_C\, cos(\omega_c t) \quad (65)$$

The flux imposed on the meminductor is given by

$$\varphi_{in} = \int V_{in}(t)dt = \int (V_m(t) + V_C(t))dt = \frac{A_m\, sin(\omega_m t)}{\omega_m} + \frac{A_C\, sin(\omega_C t)}{\omega_C} \quad (66)$$

Also, $V_{in}(t) = A_m\, cos(\omega_m t) + A_C\, cos(\omega_c t)$ (67)

Thus, the output current of the modulator, $I_s(t)$, results as:

$$I_s(t) = V_B \frac{K}{\sqrt{2}}\left(\frac{G_{m4}}{C_1}\int V_B(t)dt - V_{ss} - 2V_{th}\right) \quad (68)$$

Now using (26), (66), and (67) in (68) and passing it through BPF

$$I_s(t) = M_1 sin\omega_c t + M_2[sin(\omega_c + \omega_m) + sin(\omega_c - \omega_m)] + M_3[sin(\omega_c + \omega_m) - sin(\omega_c - \omega_m)] \quad (69)$$

Where,

$$M_1 = -\frac{K(V_{SS}+2V_{th})A_C}{\sqrt{2}R_{X2}C_2\omega_C},\ M_2 = \frac{A_C A_m G_{m4} K}{2\sqrt{2}C_1\omega_C}\left(\frac{1}{sR_{X2}C_2}\right)^2 \text{ and } M_3 = \frac{A_C A_m G_{m4} K}{2\sqrt{2}C_1\omega_m}\left(\frac{1}{sR_{X2}C_2}\right)^2$$

It is evident that (69) is in the form of components of the upper sideband, lower sideband, and carrier of standard AM expression. In order to recover the message signal from the modulated signal, a coherent product demodulator is used. The demodulator circuit is realized with OTA as a multiplier cascaded with an OTA-based low pass filter, as shown in Fig. 21(c).

*8.2 Simulation of amplitude modulator and demodulator*

The parameters used to simulate the amplitude modulator (AM) and demodulator are given in Table 5. Fig. 20(a) shows the carrier signal ($V_C(t)$), modulating signal ($V_m(t)$), and the modulated signal, $V_S(t) = I_s(t)\,R_1$ taken at the output of current mode band pass filter in Fig. 19(b). Fig. 20(b) shows the modulated signal spectrum obtained by applying the rectangular window function present in FFT mode. Fig. 20(c) shows the recovered message signal, $V_{mr}(t)$, obtained at the output of the coherent demodulator, as shown in Fig. 19(c). It confirms that the scheme of AM proposed in this section using meminductor circuit work satisfactorily. It can further be shown that one can obtain under, over, and critical modulations by varying the amplitude of carrier and message signal. Figures 20(d) and 22(e) show amplitude-modulated waveform and its hysteresis loop for the one-time period, respectively. On analyzing Fig. 20(d) and Fig. 20(e) simultaneously, one can observe that the amplitude of the hysteresis loop formed between charge and voltage increases and decreases according to the increase and decrease in the amplitude of the message signal. So, the hysteresis loop itself changes its shape, size, and amplitude according to the message signal. For the first half cycle of the message signal, i.e., for time duration A-B in Fig. 20(d), the amplitude of the hysteresis loop in Fig. 20(e) decreases from point S to T in the positive half cycle, and point U to V in a negative half cycle of the modulated wave, simultaneously. This results in the mapping of message amplitude on the carrier signal. Similarly, for the second half cycle of the message signal, i.e., for time duration B-C in Fig. 20(d), the hysteresis loop of Fig. 20(e) increases from point T to S for the positive half cycle and point V to U for the negative half cycle of the modulated wave simultaneously leading to the mapping of message amplitude on the carrier signal. Thus, the hysteresis loop, which is nothing but a meminductance slope, is frozen as the peak amplitude variation of the carrier, i.e., it follows the amplitude of the message signal. From Fig. 20(e), we find that the hysteresis loop overlaps each other on each half-cycle.

**Table 5. AM circuit simulation parameter**

| **S.No.** | **Parameter** | **Value** |
|---|---|---|
| 1 | Message signal amplitude Am | 120 mV |
| 2 | Message signal frequency fm | 50 KHz |
| 3 | Carrier signal amplitude Ac | 370 mV |
| 4 | Carrier signal frequency fc | 1 MHz |
| 5 | Band Pass filter center frequency | 1 MHz |
| 6 | Band Pass Resistance ($R_1$=$R_2$=$R_5$) | 200 Ω |
| 7 | Band Pass Filter Capacitance($C_3$=$C_4$) | 150 pF |
| 8 | Capacitance $C_1$ | 32 pF |
| 9 | Capacitance $C_2$ | 150 pF |
| 10 | Local carrier amplitude | 450 mV |
| 11 | Local carrier frequency fc | 1 MHz |
| 12 | Low Pass filter cut-off frequency | 50 KHz |
| 13 | Low Pass Resistance R2 | 200 Ω |
| 14 | Low Pass Filter capacitance ($C_1$=$C_3$) | 10 pF |

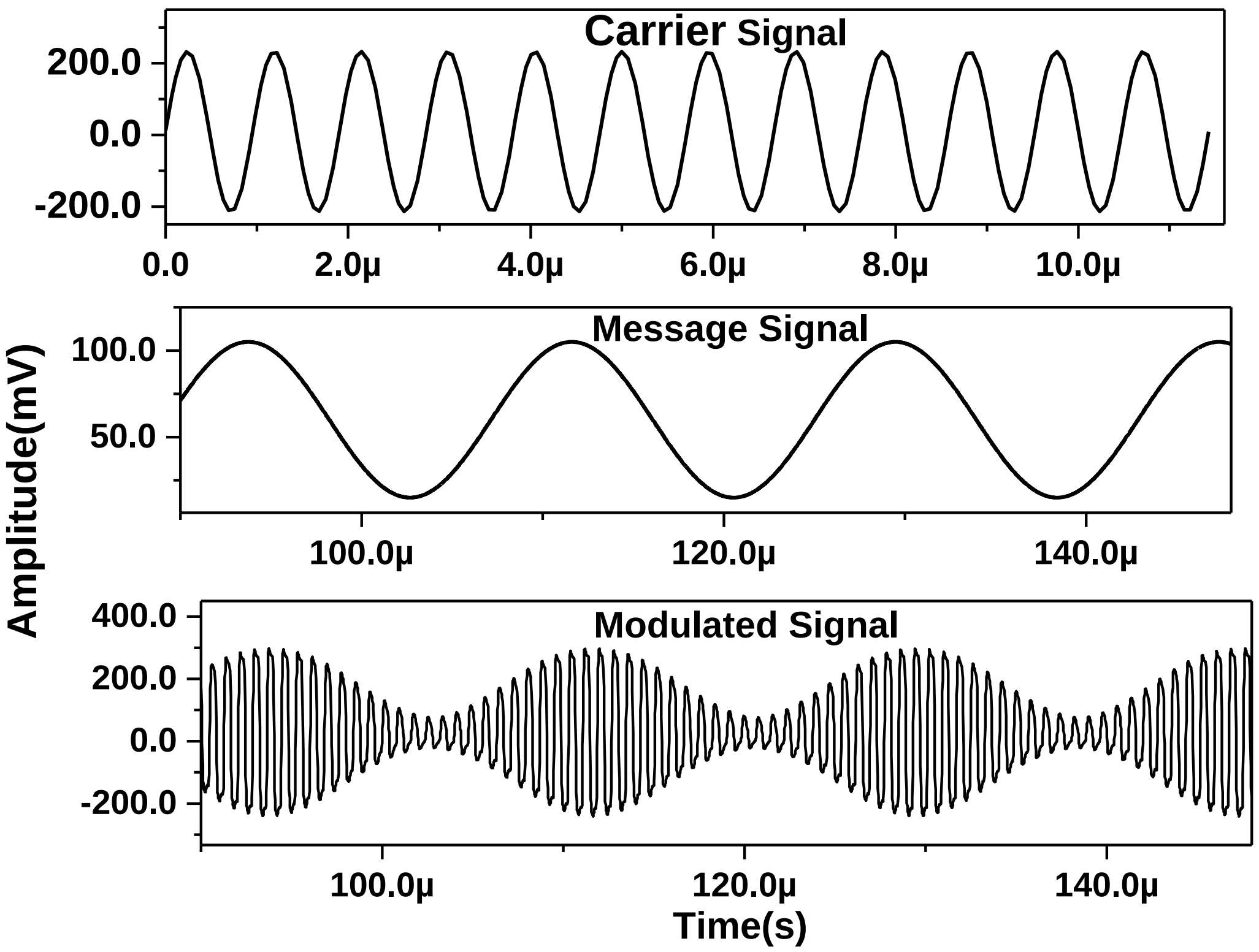

(a)

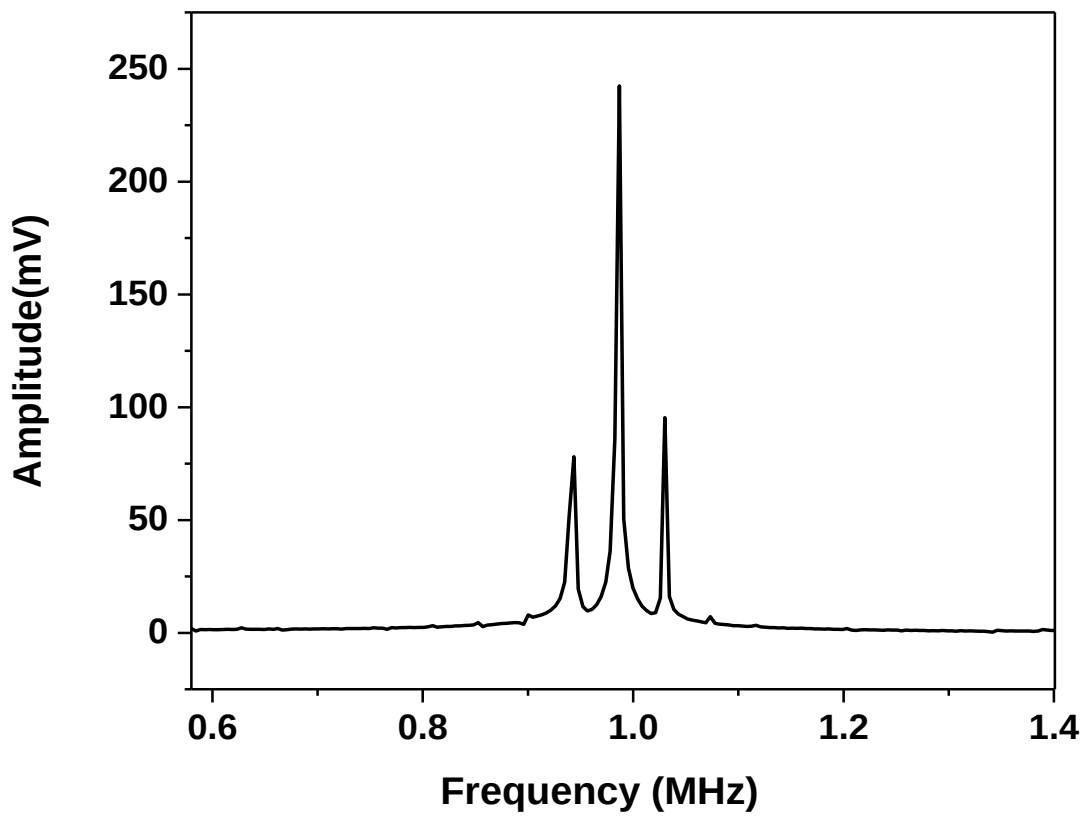

(b)

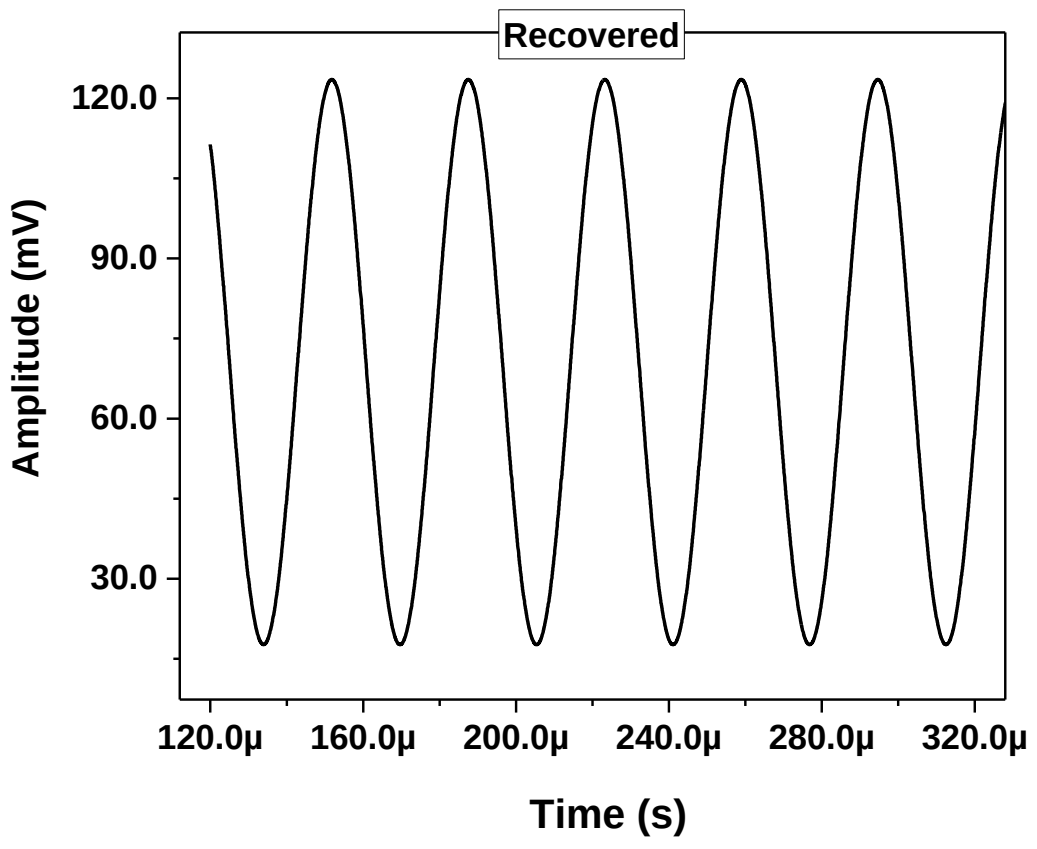

(c)

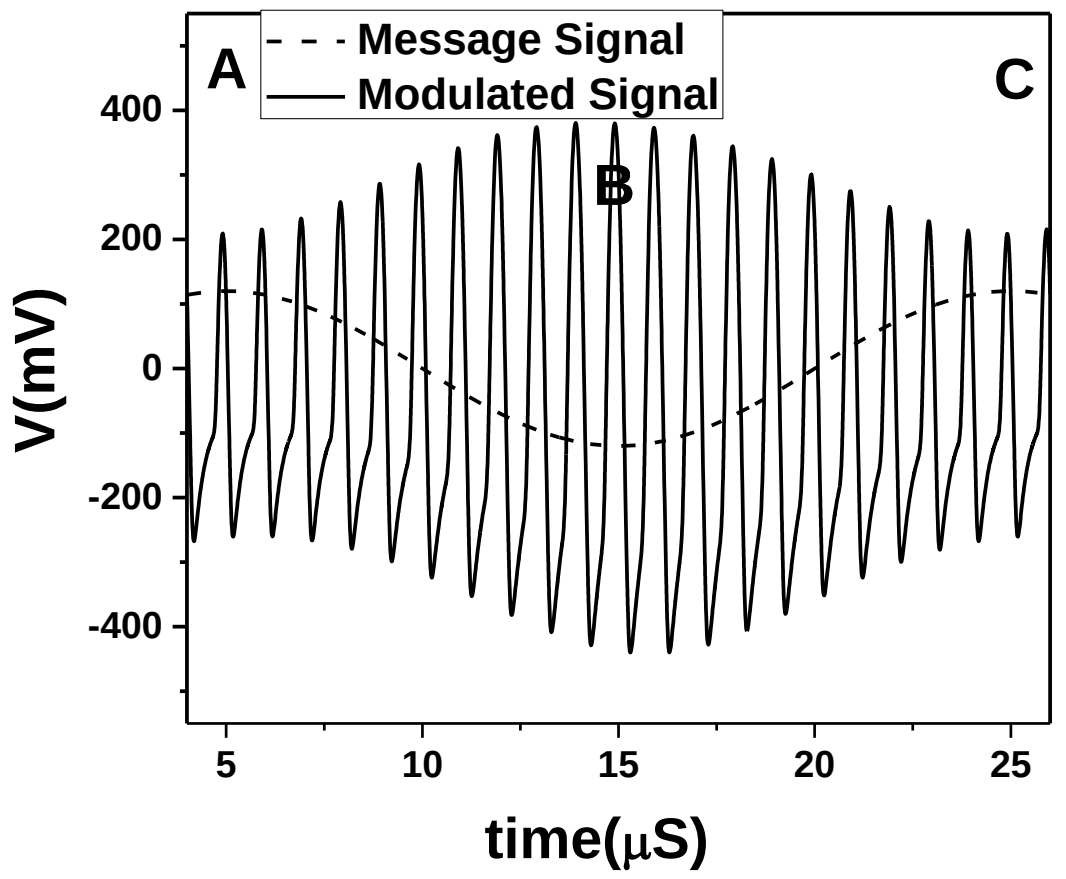

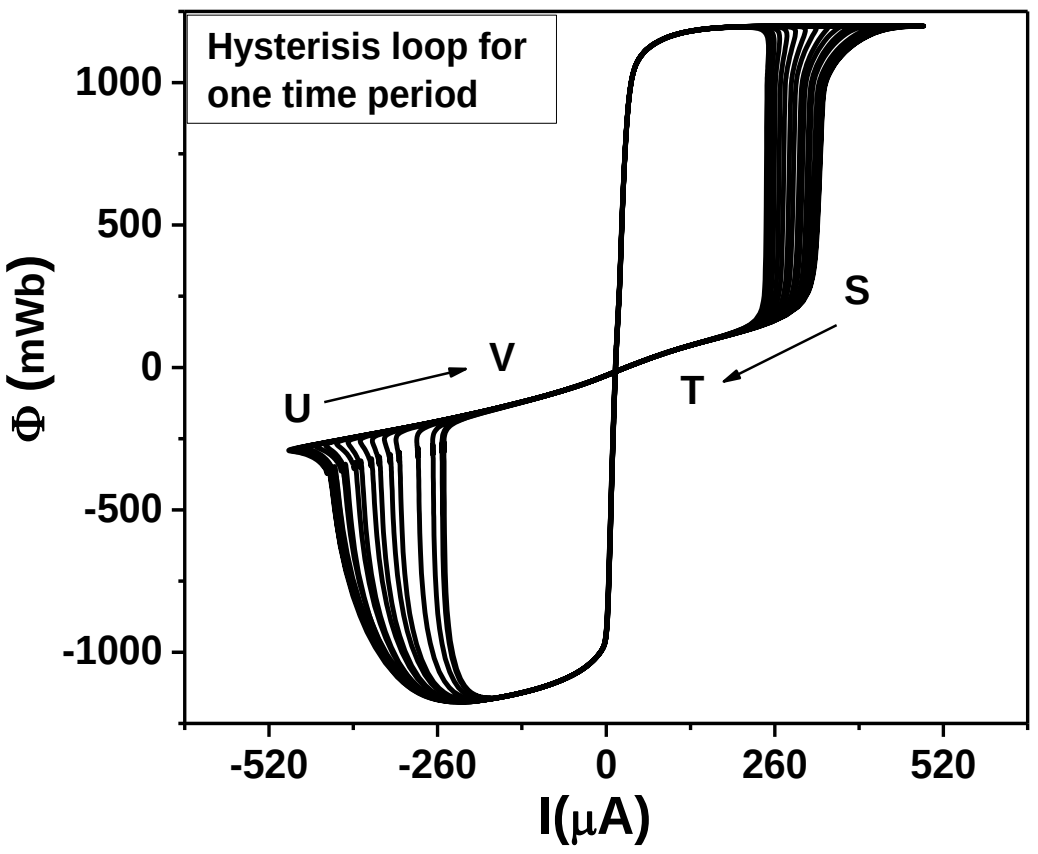

(d) (e)

**Fig. 20.** The plot of (a) carrier signal, message signal, and modulated signal, (b) spectrum of the modulated signal, (c) recovered message signal, (d) waveform of message and carrier for one time period, (e) hysteresis loop of AM for one time period.

## 9. Experimental results

As monolithic implementations of the proposed meminductor are not available, the possible experimental realization of meminductors using commercially available ICs, AD844 and CA3080, is shown in Fig. 21 (a, b), followed by the assembled grounded meminductor emulator on a breadboard is given in Fig. 21 (c). This circuit can broadly verify the functionality; however, detail parameters cannot be comparable with monolithic implementation. To verify experimentally, the proposed meminductor emulator is implemented with capacitance ($C_1$) as 40nF (in a parallel combination of four 10nF), resistance ($R_1$) of 100 Ω, and commercially available BJT-based OTA ICs (CA3080), and CFOA ICs (AD844). The pinched hysteresis loops of the emulator are obtained for the operating frequency of 820 kHz for a 4V peak-to-peak input signal, as shown in Fig. 22(a). Fig. 22(b) shows the pinched hysteresis loops for a floating meminductor (hardware implementation not shown) emulator at an operating frequency of 910 kHz for a 4V peak-to-peak input signal. Fig. 22(c) shows the time domain waveform for the grounded meminductor's current and charge (integration of current). Fig. 22(d) shows the time domain waveform for input phi and rho  (integration of phi) for the grounded meminductor. It can further be noted that both charge and rho curves tend to increase as time increases which justifies that the proposed circuit is a meminductor. It can further be noted that the integration of the input current and phi across the capacitor is performed through the inbuilt integration function in the oscilloscope to prevent any loading effect and loss of signal. Fig. 22(e) shows the time domain waveform for flux and current for the grounded meminductor. Analysis of Fig. 22(e) reveals that the proposed circuit possesses a nonlinear relationship between flux and current and is simultaneously zero, thus verifying passivity.

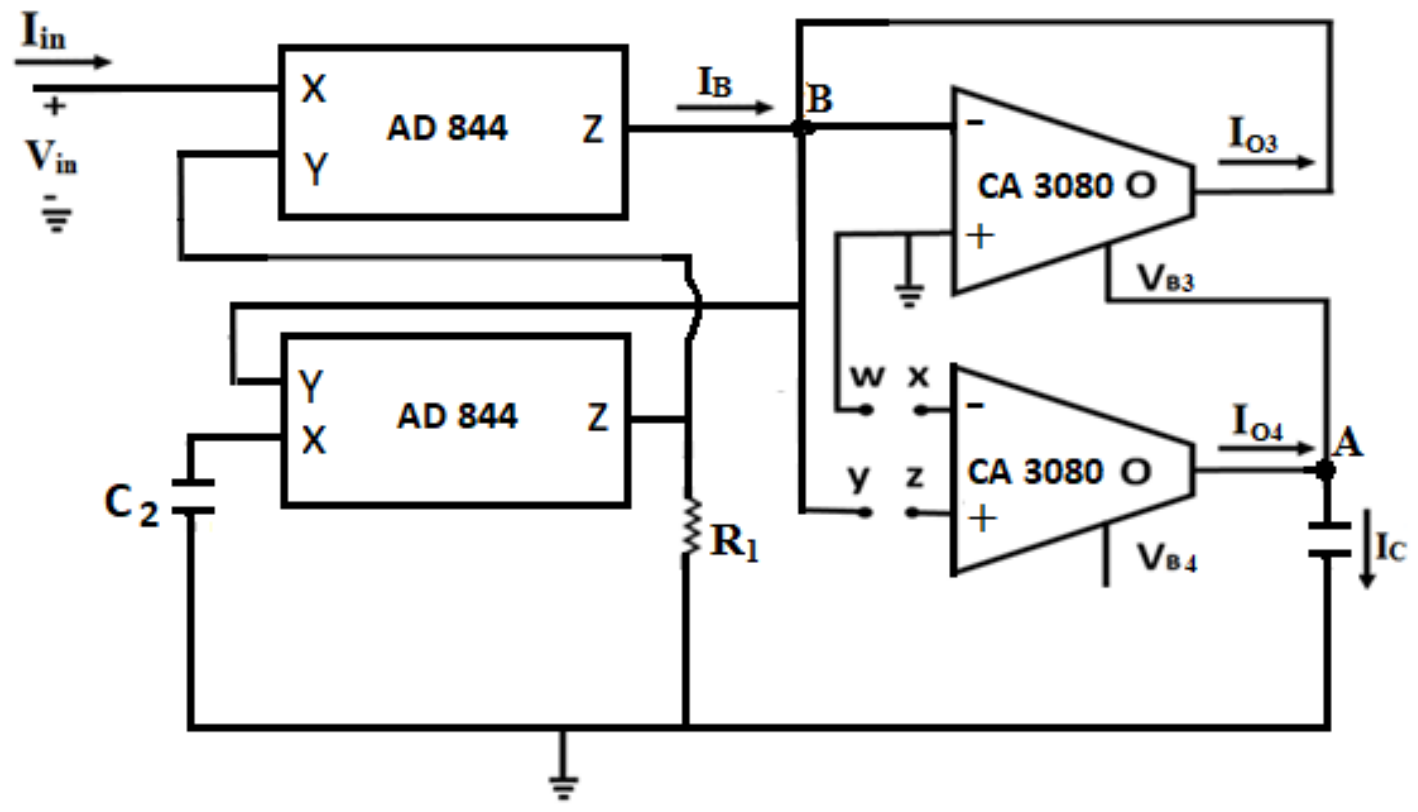

(a)

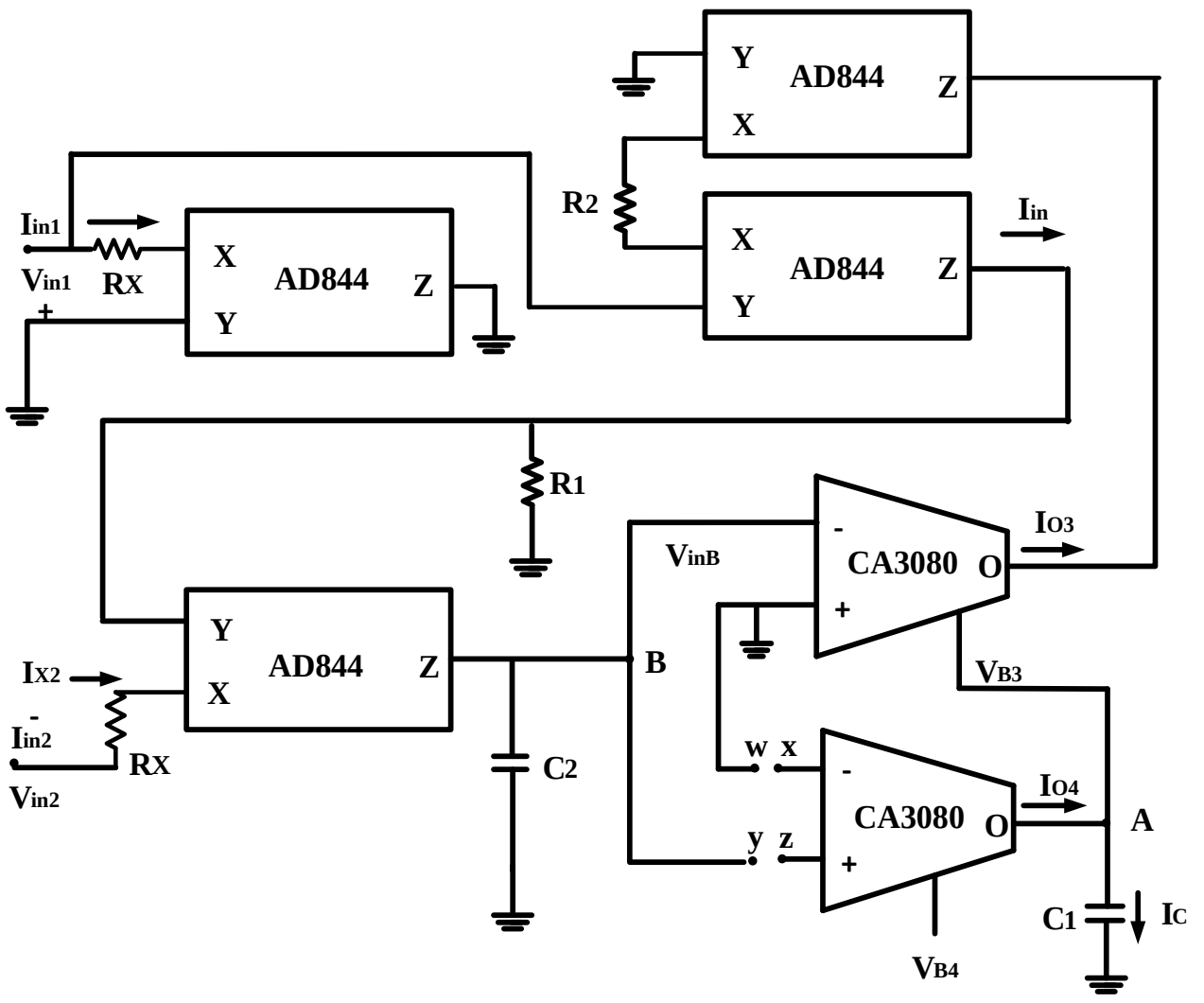

(b)

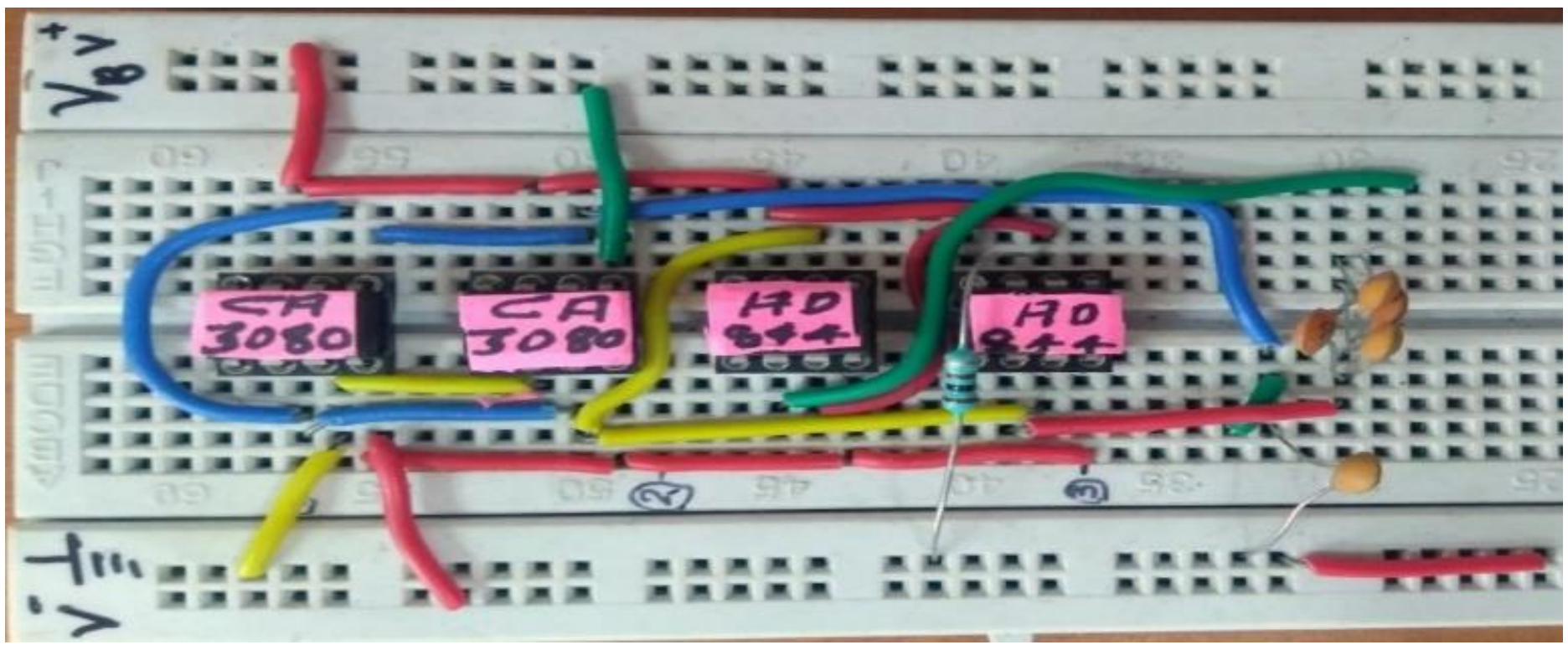

(c)

**Fig. 21.** Meminductor emulator (a) Grounded Meminductor experimental circuit, (b) Floating Meminductor experimental circuit, (c) Prototype on a breadboard

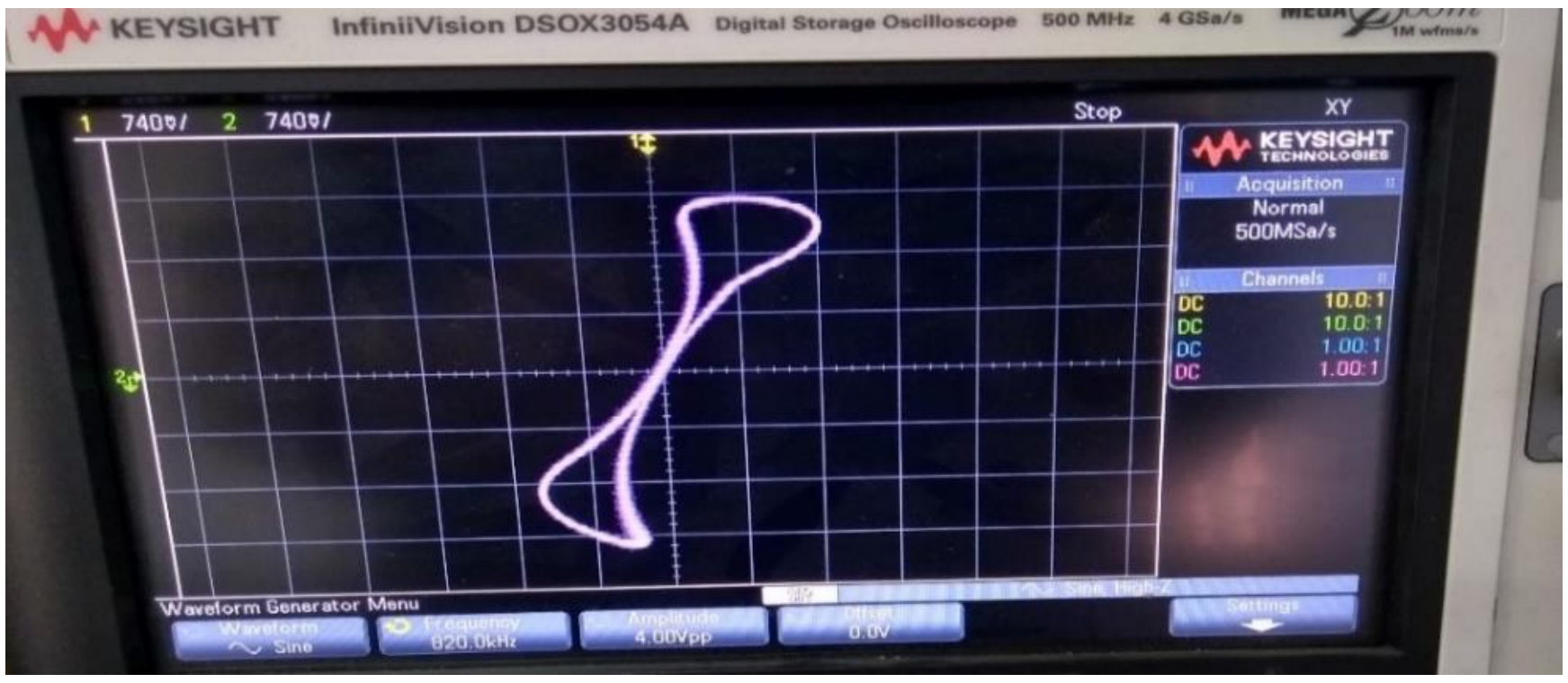

(a)

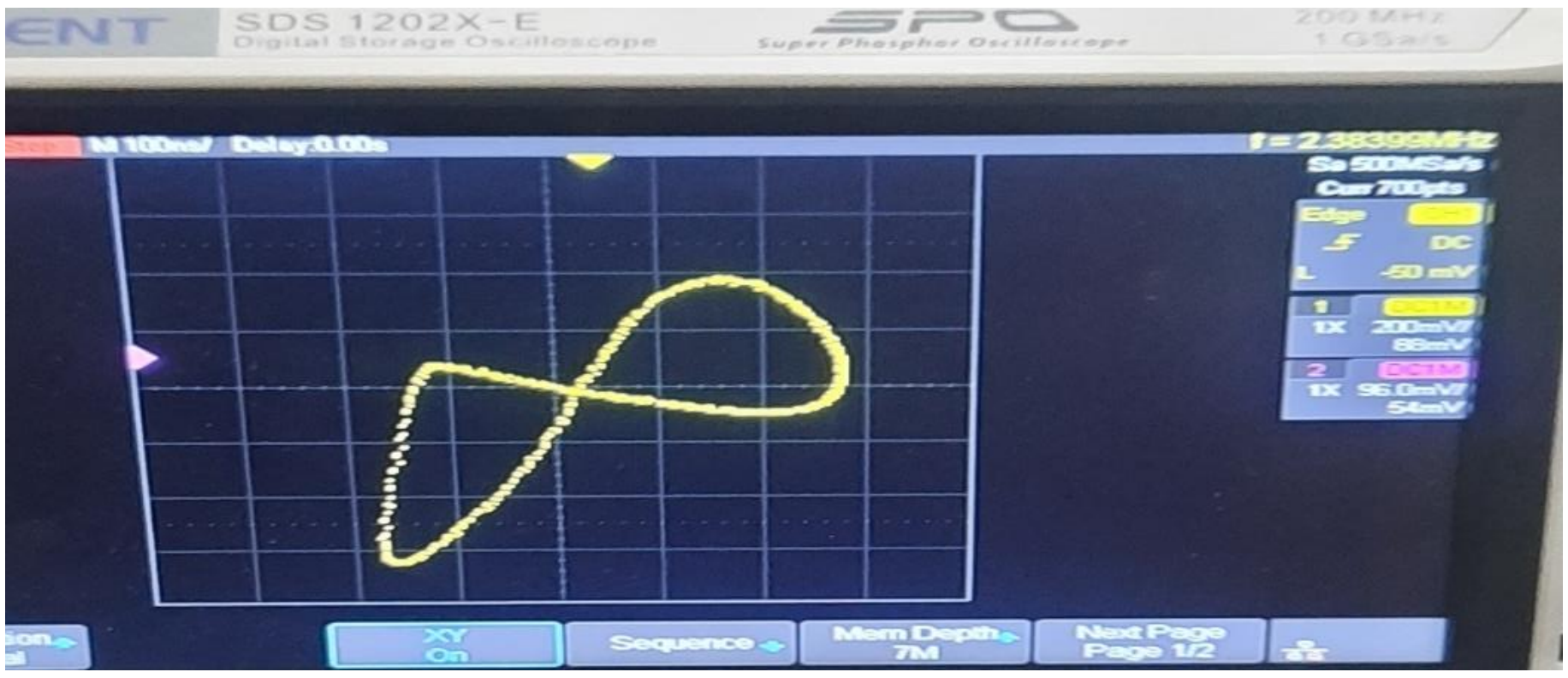

(b)

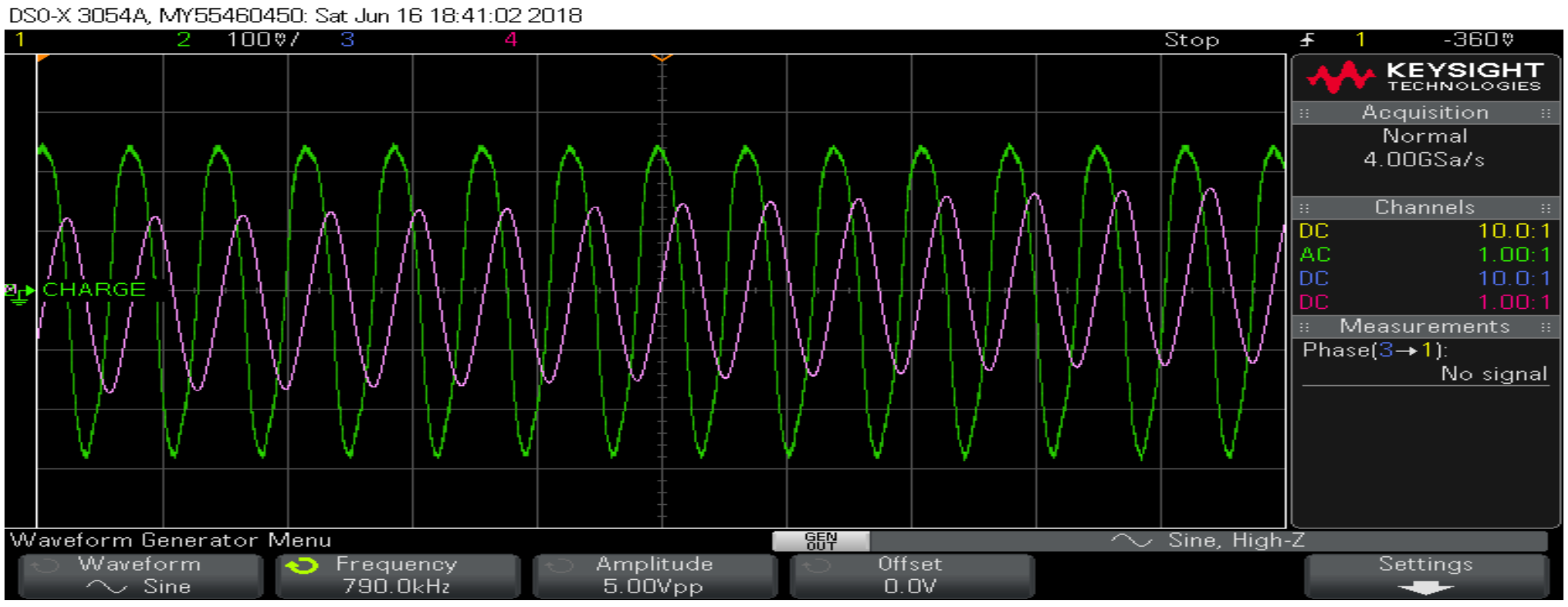
DSO-X 3054A, MY55460450: Sat Jun 16 18:41:02 2018
Stop
KEYSIGHT TECHNOLOGIES
Acquisition
Normal
4.00GSa/s
Channels
DC 10.0:1
AC 1.00:1
DC 10.0:1
DC 1.00:1
Measurements
Phase(3→1):
No signal
CHARGE
Waveform Generator Menu
Sine, High-Z
Waveform
Sine
Frequency
790.0kHz
Amplitude
5.00Vpp
Offset
0.0V
Settings

(c)

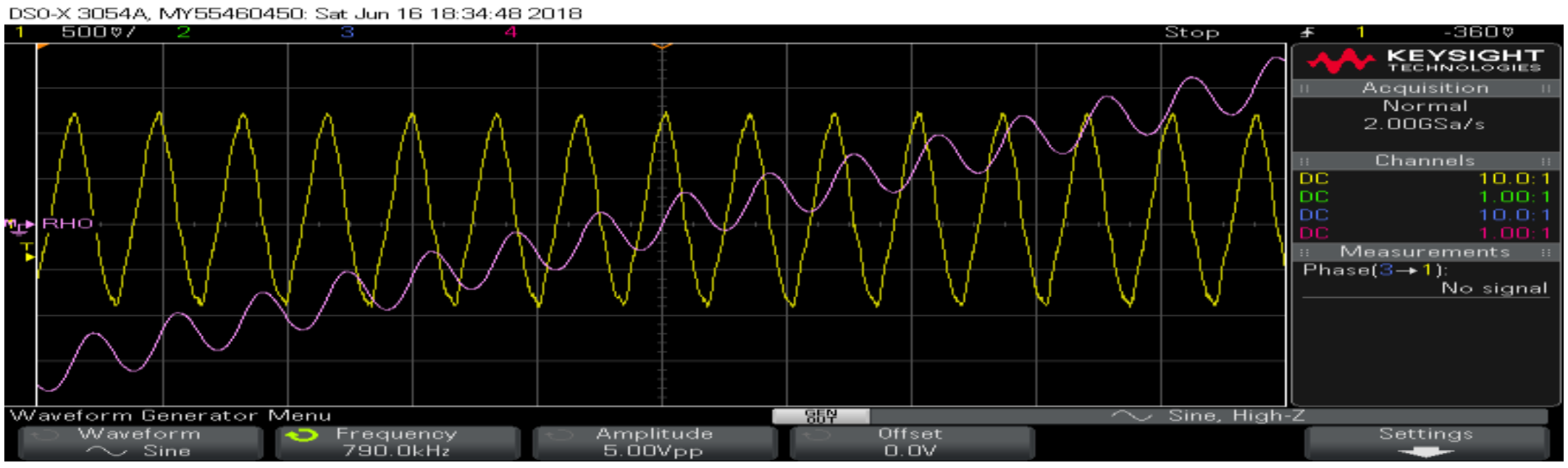
DSO-X 3054A, MY55460450: Sat Jun 16 18:34:48 2018
Stop
KEYSIGHT TECHNOLOGIES
Acquisition
Normal
2.00GSa/s
Channels
DC 10.0:1
DC 1.00:1
DC 10.0:1
DC 1.00:1
Measurements
Phase(3→1):
No signal
RHO
Waveform Generator Menu
Sine, High-Z
Waveform
Sine
Frequency
790.0kHz
Amplitude
5.00Vpp
Offset
0.0V
Settings

(d)

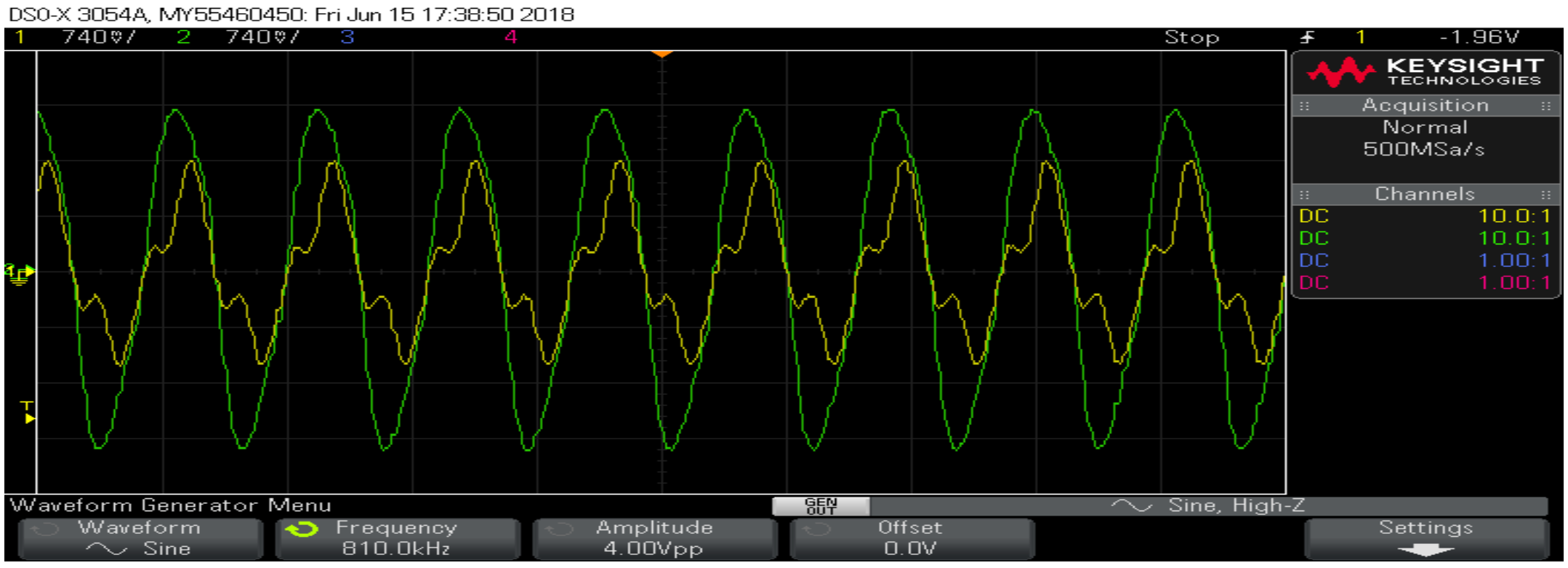

(e)

**Fig. 22.** Experimental Results of meminductor emulator (a) hysteresis loop for grounded meminductor, (b) hysteresis loop for floating meminductor, (c) time domain waveform for Current and Charge, (d) time domain waveform for phi and pho, (e) time domain waveform for phi and current. [Remove dates from Figures]

## 10. Conclusion

The proposed grounded and floating meminductor emulators show a pinched hysteresis loop and meminductive nature similar to an actual memindcutive device. Emulators have simple circuitry built with two OTAs and two CCII/CCCII. Proposed emulators are useful for a frequency of up to 1 MHz for grounded and 10 MHz for floating in both incremental and decremental topologies. The Meminductance of the proposed emulators is electronically tunable by the bias voltage of OTA and the bias current of CCCII. The controllability of the pinched hysteresis loop and the meminductance nature of proposed emulators for different frequencies of the input signal, capacitors, external bias voltage, and bias current is verified by simulation results. Layout, post-layout simulations, Monte Carlo, and detailed non-ideal analyses have also been carried out. Moreover, an AM modulator has been realized using the proposed meminductor emulator circuit as an application. Finally, a prototype of the proposed circuit is assembled, and experimental results are given and discussed.